\documentclass[singlecolumn, 
 aps,
pra,
prb,
prstab,
]{revtex4-2}

\usepackage{graphicx}
\usepackage{mathrsfs}
\usepackage{csvsimple}
\usepackage{graphicx}  
 \usepackage{amsmath}   
\usepackage{amssymb} 
\usepackage{array} 

\usepackage{bm}
\usepackage[dvipsnames]{xcolor}

\usepackage{subcaption}
\usepackage{times}
\usepackage{placeins}
\usepackage{amsfonts}
\usepackage{varioref}
\usepackage{textcomp}

\usepackage{hyperref}
\usepackage{hyperref}
\usepackage{booktabs}
 \labelformat{section}{Section #1} 
\labelformat{subsection}{Section #1} 
\labelformat{subsubsection}{Section #1}
\labelformat{subsubsubsection}{Section #1}
 \labelformat{equation}{Eq.~(#1)} 
\labelformat{figure}{Fig.~#1} 
\labelformat{subfigure}{Fig.~\thefigure#1} 
 \labelformat{table}{Table~#1} 
 \labelformat{appendix}{Appendix #1}

\begin{document}

\hbadness=99999
\preprint{APS/123-QED}
\title{Perturbations in the parametrized wormhole spacetime and their related quasinormal modes}

\author{Shauvik Biswas}
\email{shauvikbiswas2014@gmail.com}
\affiliation{Department of Physics, Indian Institute of Technology, Guwahati 781039, India}

\author{Sayan Chakrabarti}
\email{sayan.chakrabarti@iitg.ac.in}
\affiliation{Department of Physics, Indian Institute of Technology, Guwahati 781039, India}


\begin{abstract}
We study electromagnetic perturbations and the associated quasinormal modes (QNMs) of parametrized static, spherically symmetric wormhole spacetimes, focusing on Damour–Solodukhin and braneworld geometries as well as their galactic extensions. Using the Bronnikov–Konoplya–Pappas (BKP) parametrization, we express the metric functions in terms of a compactified radial coordinate and characterize the spacetime through far-field and near-throat parameters. The far-field coefficients govern the asymptotic structure and post-Newtonian behaviour, while the near-throat continued-fraction expansion captures the strong-field geometry near the throat. We first apply the parametrization to isolated wormholes and identify its range of validity, showing that the non-analytic metric functions can limit the applicability of the BKP scheme. Moreover we have shown that such issues can indeed hamper the primary ringdown waveform. We then extend the framework to a galactic Damour–Solodukhin wormhole embedded in a Hernquist dark matter halo. Imposing observational bounds from the shadow of Sgr A$^*$, we constrain the galactic compactness and deformation parameters and obtain an observationally viable parametrized metric. Within the allowed parameter space, we compute the fundamental QNM frequencies using the transfer matrix method and analyze the corresponding time-domain ringdown signals. We find that the damping rate is more sensitive to galactic compactness, whereas the oscillation frequency remains comparatively stable. Although the spectral shifts are small within the shadow-allowed region, the framework provides a systematic link between geometric parametrization, shadow constraints, and dynamical response. Our results establish an observationally consistent parametrized description of wormhole perturbations for strong-field tests of horizonless compact objects.
\end{abstract}
\maketitle

\section{Introduction}\label{sec-1}

The general theory of relativity (GR) has so far remained extraordinarily successful in describing gravitational phenomena across a wide range of scales and has passed all major experimental tests. Nevertheless, most of these tests probe only the weak-field regime. There are both theoretical and observational reasons to expect that GR may require modifications in regions of strong gravity or large spacetime curvature; see \cite{Berti:2015itd} for a detailed review. A direct way to extract information from such regions is to perturb the compact object sourcing the gravitational field and study its response. For black holes (BHs), this response is encoded in quasinormal modes (QNMs)~\cite{Cardoso:2025npr}, which dominate the ringdown phase and carry information about the underlying spacetime. In the standard BH picture, part of the perturbation generated near the light ring propagates outward and can be observed at infinity, while the rest falls through the horizon. Since the horizon behaves as a one-way membrane, that part leaves no direct observational imprint. This picture changes for horizonless exotic compact objects (ECOs) \cite{Cardoso_2019}. In the absence of an event horizon, perturbations can interact with the compact object boundary and return to distant observers, potentially carrying information from the near-surface region.

Among horizonless compact objects, wormholes (WHs) are particularly interesting. They are non-singular geometries that connect two distant regions of the same universe, or even two different universes, through a bridge-like structure \cite{Visser:1995cc}. They may also be viewed as members of the broader class of ECOs \cite{Cardoso_2019}. Originally introduced in the form of the Einstein-Rosen bridge \cite{Einstein:1935tc}, wormholes later reappeared in modern gravitational physics as potentially traversable geometries under suitable conditions \cite{Morris:1988cz}. Over the years, they have attracted sustained attention in several contexts, including modified gravity \cite{Lobo:2009ip,Harko:2013yb}, quantum gravity \cite{Maldacena:2013xja,Brown:2019hmk}, and strong-field astrophysics \cite{Sen:2024rna, Kumar:2023wfp, Kumar:2024vdh,Biswas:2026vdp}. From an observational perspective, wormholes are compelling because they can mimic BHs while differing from them in one essential aspect, viz., they do not possess an event horizon. Like other ECOs, such as gravastars and fuzzballs, they may instead exhibit reflective or partially reflective surfaces \cite{Cardoso_2019, Cardoso:2016oxy}. This difference can leave distinct signatures in their dynamical response, including deviations in the QNM spectrum and the possible appearance of gravitational wave echoes \cite{Mark_2017, Cardoso:2016oxy}. Such features make wormholes useful theoretical laboratories for identifying possible departures from the classical BH paradigm. At the same time, they may offer insight into deeper issues in gravity, including information loss and near-horizon quantum effects \cite{Maldacena_2013, Baez:2014bka}. Wormholes, however, are not free from complexities. In many constructions, exotic matter is required to keep the throat open \cite{Morris:1988cz,Cardoso:2016rao}. Many solutions are also dynamically unstable \cite{Bronnikov:2001ils, Bronnikov:2012ch, Gonzalez:2008wd, Maggio:2018ivz}. It is therefore important to identify frameworks in which their physical viability and observational signatures can be studied in a systematic way. This is because astronomical observations can not only test the existence of the horizon, but they can also put constraints on the deviation from general relativity. Since the number of alternative theories and compact-object models is very large, testing each model separately against a given observation is not practical. To overcome this issue, the authors of~\cite{Rezzolla:2014mua} formulated the idea of parametrized spacetime where a parametrization of the most general spherically symmetric black hole spacetime was performed. The main advantage of this scheme over the standard Taylor series expansion \cite{Johannsen:2011dh} is that, one can keep track on the terms which dominate in the different parts of the spacetime. With a similar motivation, the parametrization of the wormhole spacetimes was presented in \cite{Bronnikov:2021liv}. Instead of working with one exact solution at a time, one introduces a family of metrics governed by physically meaningful parameters and then analyzes the associated phenomenology in a broader setting.~Parametrized wormholes thus provide a flexible way to describe a wide class of geometries, including those that may eventually be constrained by observations  \cite{Cardoso:2016oxy,Bronnikov:2012ch}. Despite their theoretical interest, the QNM spectra of parametrized wormholes have not yet been studied in comparable detail to those of BH spacetimes. In BH physics, QNMs have long served as one of the main tools for analyzing their dynamical response and have become central to gravitational wave astronomy \cite{Berti:2009kk, Konoplya:2011qq, Berti:2025hly}. For wormholes, they offer a similarly powerful diagnostic because they are sensitive to the underlying geometry and can help distinguish WHs from BHs and other ECOs \cite{Cardoso:2016oxy}. However, while the QNM spectra of Schwarzschild, Kerr, and Reissner-Nordstr\"{o}m BHs, as well as many BH solutions in modified gravity, are now known in considerable detail, analogous results for parametrized wormholes remain much more limited \cite{PhysRevD.109.084013, PhysRevD.97.124004}. In particular, a systematic description of how the parametrization coefficients affect mode frequencies, damping timescales, and spectral structure is still lacking.


This question has become more timely in view of recent observational developments. The Event Horizon Telescope (EHT) observations of M$87^*$ and Sgr $A^*$ have opened a new avenue for constraining compact-object geometries through direct imaging and shadow measurements \cite{EventHorizonTelescope:2019dse}. These observations already place strong bounds on deviations from the classical Kerr picture, thereby restricting the viable parameter space of alternative compact-object models \cite{Cunha:2019universe, Afrin:2023tool, EHT:2022paperVI, Vagnozzi:2023tests, Khodadi:2020wake,Kumar:2023wfp, Kumar:2024vdh, Gera:2024qob}. Incorporating such constraints into the study of parametrized wormholes is therefore essential if one wishes to make realistic predictions for their QNM spectra. Future improvements in EHT observations and black hole imaging are expected to sharpen these bounds further and may eventually allow precision tests of specific wormhole scenarios \cite{Psaltis:2018xkc}.

In this work, we use the Damour-Solodukhin~(DS), braneworld, and galactic Damour-Solodukhin WHs as controlled backgrounds to examine the applicability of the BKP parametrization and its impact on dynamical observables. For the isolated DS wormhole, the BKP representation reproduces the exact metric. For the braneworld wormhole, the redshift function contains non-analytic dependence on the compact coordinate which limits the applicability of the BKP expansion. For the galactic DS wormhole, the transcendental form of the metric function similarly requires a controlled approximation. Our aim is therefore not merely to recompute known QNM spectra, but to quantify how the parametrized representation affects the light-ring structure, QNM frequencies, and primary ringdown waveform. Towards that direction, we have considered the parameter space in which the exact metric components and the parametrized metric components matches. In the literature, the scalar perturbations of these geometries were studied in detail~\cite{Biswas:2023ofz,Bueno:2017hyj}. 
In this work, we restrict our analysis to electromagnetic perturbations, since they provide a clean probe of the background geometry and lead to a comparatively simple effective potential, making the numerical computation more tractable. Although electromagnetic perturbations were studied in full generality in    ~\cite{Biswas:2022wah}, the role of the parametrized representation and its impact on the resulting spectrum were not addressed there. For the galactic Damour-Solodukhin wormhole, we further test the parametrization against Sgr~A$^{*}$ shadow constraints using two complementary approaches: first, by deriving the allowed galactic compactness from the exact shadow expression and checking the consistency of the BKP parameters, and second, by applying the shadow constraint directly to the parametrized metric. Together with the existing bound on $\lambda$~\cite{Karimov:2019qfw}, this provides an observationally consistent parameter space for studying how the BKP parametrization affects the QNM spectrum and ringdown waveform. To the best of our knowledge, this work provides one of the first attempts toward the discussion of the applicability of the BKP scheme throughout the WH spacetime and thereby exploring the effect of parametrization on the QNMs as well as on the ringdown waveform.



The paper is organized as follows. In \ref{sec2}, we review the parametrization of static, spherically symmetric wormhole spacetimes in the Morris--Thorne frame. In
\ref{sec-3}, we demonstrate the parametrization for isolated Damour-Solodukhin and braneworld wormholes. \ref{sec-4-0} presents the formulation of electromagnetic perturbations and the computation of quasinormal modes. In \ref{sec-5}, we construct the parametrized galactic Damour-Solodukhin wormhole, followed by consistency checks with shadow observations in 
\ref{sec-6}. The quasinormal ringing of the observationally consistent parametrized wormhole is analyzed in \ref{sec-7}. In \ref{sec-8}, we have further explored the sensitivity of the primary signal on the parametrization.
Finally, in \ref{sec-9}, We conclude with a discussion of the implications of our results and possible extensions of this work.
In this paper, we will work with positive signature metric and set $c=G=1$.

\section{Parametrization of Static Spherically Symmetric Wormhole Spacetime in the Morris-Thorne (MT) frame}\label{sec2}

In this section, we will briefly review the Bronnikov-Konoplya-Pappas (BKP) parameterization \cite{Bronnikov:2021liv} for any static spherically symmetric asymptotically flat WH spacetime. The parametrization provides a hierarchical representation~\cite{Rezzolla:2014mua} of the metric functions in terms of a compactified radial coordinate, enabling a systematic separation between near throat and asymptotic (far field) behavior. To start with, we assume that the WH spacetime is described by the following line element,
\begin{align}\label{General-static-metric}
    ds^{2}=-f(r)dt^{2}+\frac{dr^{2}}{h(r)}+K(r)^{2}d\Omega_{2}^{2}~.
\end{align}
Here, $K(r)$ can be thought of as the circumferential radius of the 2-sphere at radial location $r$. In our case, we will take $r$ as curvature coordinate, that is $r=K(r)$ and express the metric \ref{General-static-metric} in the Morris-Thorne form\cite{Morris:1988cz}. Explicitly, this is given by,
\begin{align}\label{MT-form}
   ds^{2}=-e^{2\phi(r)}dt^{2}+\frac{ d r^{2}}{1-\frac{b(r)}{r}}+r^{2}d\Omega_{2}^{2}~,
\end{align}
here, $\phi(r)$ denotes the redshift function, while $b(r)$ represents the shape function. The utility of using MT frame lies in the fact that both Damour-Solodukhin (DS) and braneworld WHs are usually casted in this form. To parameterize the spacetime, i.e. to construct a uniformly convergent expansion over the physical domain $r\in [r_0,\infty)$, we introduce the dimensionless compact coordinate (DCC) as \cite{Rezzolla:2014mua},
\begin{align}
    x\equiv 1-\frac{r_{0}}{r}~,
\end{align}
where $r_{0}$ refers to the location of the wormhole throat and is given by the root of equation $r_{0}=b(r_{0})$, while satisfying condition $b'(r_0)<1$ \cite{Visser:1995cc}. This maps the exterior region onto the closed interval $x\in[0,1]$. In terms of $x$, the metric functions are recast as
\begin{align}
    A(x)\equiv f(r),
    B(x)\equiv h(r)~.
\end{align}
The functions $A(x)$ and $B(x)$ are decomposed into asymptotic contributions fixed by flatness and a continued-fraction expansion encoding deviations near the throat. The asymptotic sector 
enforces the correct large $r$ behavior, fixing the ADM mass and leading post-Newtonian terms. The near throat geometry, on the other hand, is represented as continued fractions \cite{Bronnikov:2021liv,Maharana:2025lbm}. Keeping the above in mind, these two functions are defined as,
\begin{align}\label{A}
    A(x)=f_{0}+x\left[(1-f_{0})-(\epsilon+f_{0})(1-x)+(a_{0}-\epsilon-f_{0})(1-x)^{2}+\frac{a
_{1}(1-x)^{3}}{1+\frac{a_{2}x}{1+\frac{a_{3}x}{...}}}\right]~,
\end{align}
\begin{align}\label{B}
    B(x)=h_{0}+x\left[(1-h_{0})-(b_{0}+h_{0})(1-x)+\frac{b
_{1}(1-x)^{2}}{1+\frac{b_{2}x}{1+\frac{b_{3}x}{...}}}\right]~.
\end{align}
The asymptotic behavior of these two functions can be found by expanding them around $x=1$ (that is, around spatial infinity), such that
\begin{align}
    &A(x)=1-(1+\epsilon)(1-x)+a_{0}(1-x)^{2}+\mathcal{O}((1-x)^{3})\label{a0}~\\
    &\frac{1}{B(x)}=1+(1+b_{0})(1-x)+\mathcal{O}((1-x)^{2})\label{b0}~.
\end{align}
From these expansions, it is clear that the three parameters, $\epsilon$,~$a_{0}$ and $b_{0}$ determine asymptotic behaviour of the metric. We will call them far field BKP parameters. One can read off these parameters by expanding $f(r)$ and $(h(r))^{-1}$ around spatial infinity and then comparing with \ref{a0} and \ref{b0}. From the above expansion of $A(x)$ around $x=1$, one can obtain the following hierarchy of the parameters {\footnote{One should always check this hierarchy for a valid parametrization.}},

\begin{align}\label{hierarchy-parameters}
|a_{0}|\leq |(1+\epsilon)|~.    
\end{align}

After determining these parameters one can find the near field BKP parameters $(a_{i},b_{i})$ with $i\neq 0$ by expanding these parameterization functions around $x=0$ (around the throat of the WH) and plugging the values of $a_{0}$, $b_{0}$ and $\epsilon$. Note that while determining $a_{i}$, one must set $a_{i+1}=0$, and similarly for $b_{i}$'s. In particular, if we keep terms up to $a_{1}$, then the expansion of $A(x)$ around $x=0$ leads to,
\begin{align}\label{A-o(x)}
    A(x)=f_{0}+(1+a_{0}+a_{1}-3f_{0}-2\epsilon) x+ \mathcal{O}(x^{2})~.
\end{align}
From a knowledge of $a_{0}$, $f_{0}$ and $\epsilon$ one can determine $a_{1}$ by expanding $f(r)$ around $r=r_{0}$ and keeping terms up to linear order in the expansion. Note that the parametrization therefore admits a key feature: truncation at a given order preserves global regularity while systematically improving accuracy both near the throat and at infinity. This is particularly advantageous for WH metrics containing non-polynomial functions, where naive Taylor expansions\cite{Johannsen:2011dh} around $r_0$ exhibit poor convergence.
We apply this construction to the Damour–Solodukhin and braneworld wormholes. For each case, the exact metric functions are expanded in $x$, and the parametrized coefficients are obtained through order-by-order matching. From a phenomenological viewpoint, it can be demonstrated\cite{Konoplya:2022tvv} that only a small number of coefficients will suffice in the continued fraction expansion. In contrast, simple polynomial truncations show degraded performance when non-analytic structures appear in the redshift function, particularly near the throat. This analysis establishes that the BKP hierarchy provides a stable and systematically improvable reconstruction of wormhole geometries suitable for perturbative studies. The parametrized metric preserves regularity at the throat and asymptotic flatness at infinity by construction, thereby ensuring a well-defined background for QNM calculations.

\section{Demonstrating the parametrization with Isolated wormholes}\label{sec-3}
\subsection{Damour-Solodukhin Wormhole}
The Damour-Solodukhin spacetime \cite{Damour:2007ap} is described by the following line element,
\begin{align}\label{DS}
    ds^{2}=-\left(1-\frac{2M_{1}}{r}\right)dt^{2}+\frac{dr^{2}}{\left(1-\frac{2M_{2}}{r}\right)}+r^{2}d\Omega_{2}^{2},~~~\text{with}~M_{2}=M_{1}(1+\lambda^{2})~.
\end{align}
In order to mimic the results of Schwarzschild spacetime, one usually takes the dimensionless parameter $\lambda\ll 1$. 
One can easily check that, $r=2M_{2}$ is not an horizon as $g_{tt}\neq 0$ at $r=2M_{2}$. Moreover, for $2M_{1}<r<2M_{2}$, the metric has a wrong signature. Therefore, one considers two copies of such space-time and glues them across $r=2M_{2}$. This construction gives the Damour-Solodukhin wormhole, with $r=2M_2$
 representing the throat. By computing the components of the Einstein tensor associated with ~\ref{DS}, one can show that the spacetime can be thought of as sourced by an anisotropic fluid that violates the energy conditions, namely the null and weak energy conditions. This is a basic pathology of wormhole spacetime in the context of general relativity. The interested readers are referred to \cite{Visser:1995cc} for more details on this. In what follows, we will take this line element~~\ref{DS} to demonstrate the previously alluded procedure. Using the method discussed (see \ref{sec2})~above, the non vanishing parameters for this spacetime turn out to be
$f_{0}=-\epsilon$ with $\lambda^{2}+1=\frac{1}{1+\epsilon}$, implying that $\epsilon$ basically controls $\lambda$. Since $\lambda^{2}$ is positive, it turns out that $\epsilon$ must be negative with $|\epsilon|<1$. 
\subsection{Braneworld Wormhole}
\begin{figure}[h!]
\begin{subfigure}{0.45\textwidth}
 \includegraphics[width=0.9\linewidth]{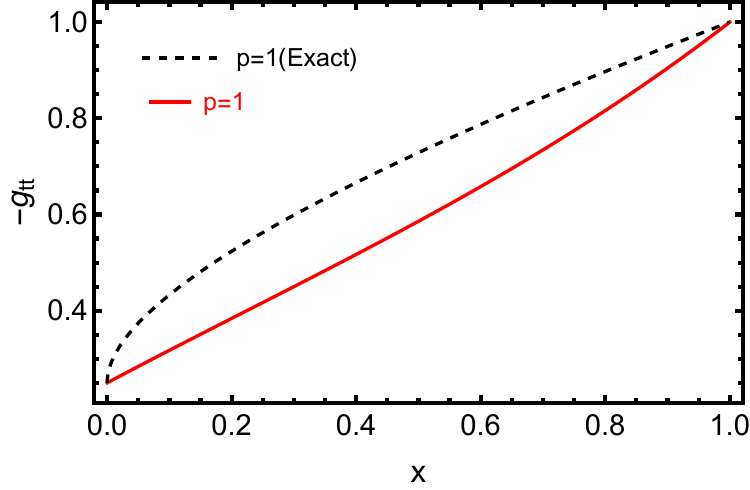}
\end{subfigure}
\begin{subfigure}{0.45\textwidth}
\includegraphics[width=0.9\linewidth]{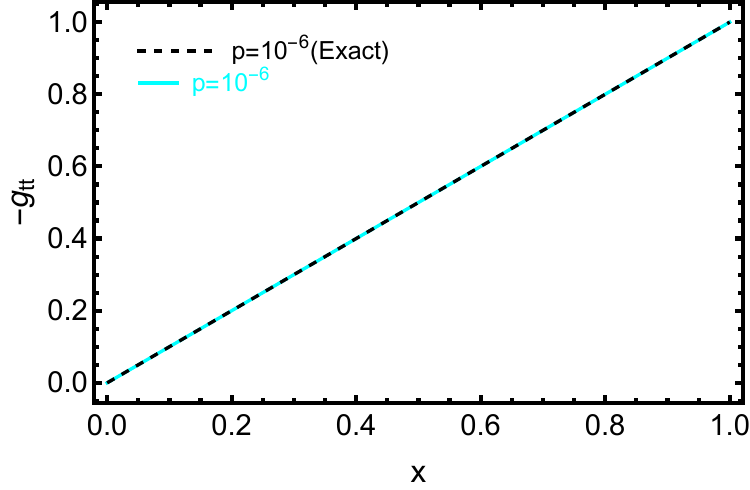}    
\end{subfigure}
\begin{subfigure}{0.45\textwidth}
    \includegraphics[width=0.9\linewidth]{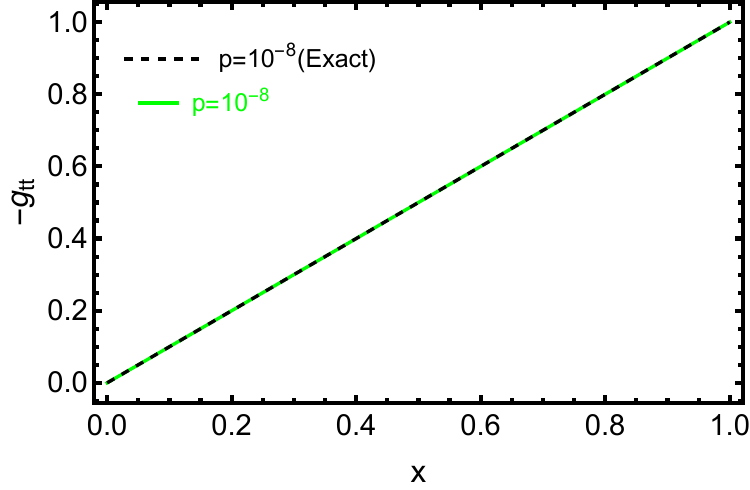}
\end{subfigure}
\caption{Comparison of the time-time component of the parametrized metric with that of exact metric components in case of the braneworld WH. Black dashed curve shows the plot for exact metric components with different values of $p$. From these figures, it is clear that as the value of $p$ increases, the parametrization do not match with that of the exact metric. This is consistent with the fact that for large values of $p$, the parametrization of the braneworld metric as given in the text becomes poor because one cannot incorporate the effect of near field parameters $a_{i}$'s in the parametrization.}\label{braneworld-metric-comparison}
\end{figure}

Having demonstrated the procedure of parametrization, we will now apply it to the wormholes on the brane. In the braneworld scenario, the four-dimensional visible universe is described as a brane embedded in a higher-dimensional bulk. Considering the curved RS1 braneworld scenario and following the low energy expansion scheme proposed by Kanno and Soda~\cite{Kanno:2002ia}, one can show that the effective Einstein's equation on the visible brane reads (assuming the Planck brane to be vacuum),
\begin{align}\label{eom-B}
G_{\mu\nu}=\frac{\kappa^{2}}{\ell \Phi}T^{\rm B}_{\mu\nu}+\frac{1}{\Phi}T^{\Phi}_{\mu\nu}~.
\end{align}
Here $\kappa$ denotes the $5\rm D$ Newton's constant and $T^{B}_{\mu\nu}$ denotes the energy momentum tensor of the matter trapped on the visible brane. On the other hand $T^{\Phi}_{\mu\nu}$ denotes a tensorial object made out of derivatives of $\Phi$. Due to it's complicated expression, we do not write it here (see\cite{Kanno:2002ia} ). 
Here $\ell$ denotes the size of the extra spatial dimension and $\Phi$ is the radion incarnation on the visible brane, which is related to the bulk radion field $\phi$ as $\Phi=\exp[2e^{\phi(x)}]-1$. Assuming that the matter energy momentum tensor on the visible brane is traceless and is that of an anisotropic fluid, one can show that the radion incarnation $\Phi$ satisfies the following equation on the visible brane,
\begin{align}\label{EOM-Phi}
\nabla^{\alpha}\nabla_{\alpha}\Phi=-\frac{1}{2\omega+3}\frac{d\omega}{d\Phi}(\nabla^{\alpha}\Phi)(\nabla_{\alpha}\Phi)~\quad\text{with}\quad\omega=-\frac{3\Phi}{2(1+\Phi)}~,
\end{align}
The static and spherically symmetric wormhole solution of the above equations~~(\ref{eom-B} and~~\ref{EOM-Phi}) were first time proposed in \cite{Kar:2015lma}. The braneworld wormhole is described by the following line element,
\begin{align}
    ds^{2}=-\frac{1}{(p+1)^{2}}\left(p+\sqrt{1-\frac{2M}{r}}\right)^{2}dt^{2}+\frac{dr^{2}}{1-\frac{2M}{r}}+r^{2}d\Omega_{2}^{2}\quad\text{with}\quad p\gtrsim 0.
\end{align}
 With similar lines of argument as given for the Damour-Solodukhin wormhole, it can be seen that the throat is located at $r=2M$. That is, in this, case the parametrized metric function $B(x)$ is just $x$. In this case, expansion of $f(r)$ around $x=0$ turns out to be,
\begin{align}
    f\sim \frac{p^2}{(p+1)^2}+\frac{2 p \sqrt{x}}{(p+1)^2}+\frac{x}{(p+1)^2}+\mathcal{O}(x^{\frac{3}{2}})~, 
\end{align}
because of this non-analytic dependence on $x$, we cannot use BKP parametrization in the near throat region unless we set all $a_{i}=0$ with $i=1,2,...$. But we can use this parametrization in the far region. This gives,
\begin{align}\label{brane-param}
    f_{0}=\frac{p^{2}}{(1+p)^{2}},~\epsilon=-\frac{p}{(1+p)}, a_{0}=-\frac{p}{4(1+p)^{2}}~. 
\end{align}
This completes the parametrization of the braneworld wormhole (only using far field parameters!), and it shows that for non-polynomial metric components, the wormhole spacetime cannot be completely parametrized using the above procedure. This in turn means that we cannot rely on this parametrization for arbitrary values of the parameter $p$. To see where the above parametrization works well, we have compared the $g_{tt}$ components of the metric for various values of $p$ in \ref{braneworld-metric-comparison}. From the figure, it is clear that for $p\gtrsim 0$ the parametrization matches well with the exact metric, which is indeed the case of our interest. In the rest of the paper, we will mainly work in this regime. One can intuitively understand why for the large values of $p$, the parametrization using only far parameters will not going to work. For large values of $p$ even if the spacetime is asymptotically Minkowskian, but the region near the throat is drastically different from that of the case $p\gtrsim 0$. Moreover, from \ref{brane-param} we see that, since both the far field parameters are functions of $p$, we can take only one of them as independent. In what follows, we will take $\epsilon$ as an independent parameter.
\section{Electromagnetic perturbation}\label{sec-4-0}
Since our background spacetime is free of any electric charge, the electromagnetic perturbation of the background spacetime is given by,
\begin{align}\label{Maxwell}
    \nabla_{\nu}F^{\mu\nu}=0~,
\end{align}
where $F_{\mu\nu}=2\nabla_{[\mu}A_{\nu]}$ is the electromagnetic field tensor, and $A^{\mu}$ denotes the perturbing electromagnetic field. Using the anti-symmetry property of $F_{\mu\nu}$, we can show that~~\ref{Maxwell}~~reduces to
\begin{align}\label{Simplified-Maxwell}
    \frac{1}{\sqrt{-g}}\partial_{\mu}\left[\sqrt{-g}F^{\mu\nu}\right]=0~.
\end{align}
Since the background spacetime is static and spherically symmetric, this equation can be decomposed into radial and angular parts by using vector spherical harmonics. Moreover, as the spherical symmetry preserves parity, so by the use of vector spherical harmonics, we can decompose the perturbations into axial and polar sectors. The angular part carries no information about the dynamics of the perturbation, whereas the radial part does. It can be shown that the radial part in both the axial and polar sectors reduces to the following time independent Schr\"odinger like equation,
\begin{align}\label{EM-master}
    \frac{d^{2}\Psi_{\ell m}}{dr_{*}^{2}}+\left[\omega^{2}-V_{\ell}(r)\right]\Psi_{\ell m}=0~.
\end{align}
Here, the tortoise coordinate $r_{*}$ is defined as~\cite{Aneesh:2018hlp},
\begin{align}\label{tortoise coordinate}
   dr_{*}\equiv\frac{dr}{\sqrt{f(r)h(r)}}~,
\end{align}
such that $r_{*}(r_{0})=0$ and we will use $\pm r_{*}$ to cover both the universes connected by the throat. Interestingly enough, the perturbing potential takes the same form in both cases and is given by,
\begin{align}\label{EM-pot}
    V_{\ell}(r)=f(r)\frac{\ell(\ell+1)}{r^{2}}~.
\end{align}
In what follows, we will be using \ref{EM-master} for our analysis. As one of the first studies towards the perturbations of the parametrized wormhole, we restrict our analysis to the electromagnetic perturbations because they provide a clean probe of the background geometry without requiring a model-dependent treatment of matter perturbations. On the other hand, gravitational perturbations, especially in the polar sector, require a consistent parametrization of the supporting matter sector and are left for future work.
\subsection{Transfer Matrix Method}\label{sec-4-1}
\begin{figure}
    \centering
    \includegraphics[width=0.5\linewidth]{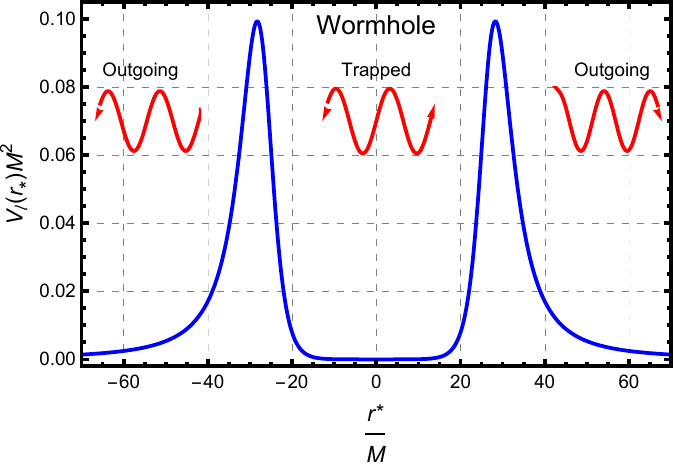}
    \caption{Schematic representation of the wormhole double bump potential $V_{\ell}(r_{*})$(see~~\ref{double-bump-potential}). The quasinormal mode problem is characterized by imposing outgoing boundary condition at both the universes. However there will be trapped modes between the bumps.}
    \label{fig:double-bump}
\end{figure}

In this method, one solves the scattering problem associated with the wormhole double bump potential by studying it with that of each single bump potential\cite{Bueno:2017hyj}. To begin with, one decomposes the double bump potential into two single bump potentials,
\begin{align}\label{double-bump-potential}
    V_{\ell}(r_{*})=\theta\left(r_{*}-\frac{L}{2}\right)V^{s}_{\ell}(r_{*})+\theta\left(-r_{*}-\frac{L}{2}\right)V^{s}_{\ell}(-r_{*})~.
\end{align}
Here, $L$ denotes the throat length, which is the measure of the distance between the peak of the double bump potential, namely $L=2|r^{\rm max}_{*}|$, such that $r_{*}^{\rm max}$ denotes the location of the maxima of the potential $V_{\ell}(r_{*})$. In~~\ref{double-bump-potential} $V^{s}_{\ell}(r_{*})$ denotes the single bump potential which appears in the following differential equation,
\begin{align}\label{single-bump-master}
   \frac{d^{2}\psi^{s}_{\ell m}}{dr_{*}^{2}}+\left[\omega^{2}-V^{s}_{\ell}(r_{*})\right]\psi^{s}_{\ell m}=0~,
\end{align}
where $\psi^{s}_{\ell m}(r_{*})$ denotes the corresponding master variable associated with each single bump potential. Since $V^{s}_{\ell}(r_{*})$ vanishes near the throat and near spatial infinity (see~~\ref{fig:double-bump}), so that $\psi^{s}_{\ell m}(r_{*})$ will have the following behaviour,
\begin{align}\
 \psi^{s}_{\ell m}(r_{*})=\begin{cases}
     A_{1}e^{-i\omega r_{*}}+A_{2}e^{i\omega r_{*}} ~\text{for}~r_{*}\rightarrow \infty
     \\\\
     B_{1}e^{-i\omega r_{*}}+B_{2}e^{i\omega r_{*}}~~\text{for}~r_{*}\rightarrow 0.
 \end{cases}
\end{align}
Now, we introduce the transfer matrix $T$ whose sole job is to relate the far amplitude $(A_{1},A_{2})$ to that of the near throat amplitudes $(B_{1},B_{2})$ in the following way~\cite{Ianniccari:2024ysv},
\begin{align}\label{Transfer-Matrix}
    \begin{pmatrix}
        A_{1}\\
        A_{2}
    \end{pmatrix}=T\begin{pmatrix}
        B_{1}\\
        B_{2}
    \end{pmatrix}~.
\end{align}
Then, due to the reflection symmetry of the problem, it follows that the transfer matrix associated with $V^{s}_{\ell}(-r_{*})$ will be $T^{\prime}=\sigma_{x}T^{-1}\sigma_{x}$, where $\sigma_{x}$ is the Pauli matrix. Further, by assuming that the master variable $\Psi_{\ell m}$ (see~~\ref{EM-master}) is continuous and differentiable across the throat, we get the total transfer matrix associated with~~\ref{double-bump-potential} to be
\begin{align}\label{t-whole}
    \mathbb{T}=T\begin{pmatrix}
        e^{i\omega L}&0\\
        0&e^{-i\omega L}
    \end{pmatrix} T^{\prime}~.
\end{align}
Now quasinormal modes of the wormhole spacetime are characterized by the fact that there are no incoming waves from $r_{*}=\pm\infty$, which, using~\ref{t-whole}, implies that $\mathbb{T}_{22}=0$ or
\begin{align}
    e^{-i\omega_{n} L}=-e^{in\pi}R_{\rm bump}(\omega_{n})~\text{for}~n=1,2,3,...
\end{align}
where $R_{\rm bump}(\omega)\equiv-\frac{T_{21}}{T_{22}}$ is the reflection coefficient of the single bump $V^{s}_{\ell}(r_{*})$ for a wave coming from near the throat region ($r_{*}\rightarrow 0$).

\subsection{Quasinormal ringing of the parametrized isolated wormholes}\label{sec-4-2}
\begin{figure}[h!]
\begin{subfigure}{0.45\textwidth}
\includegraphics[width=0.9\linewidth]{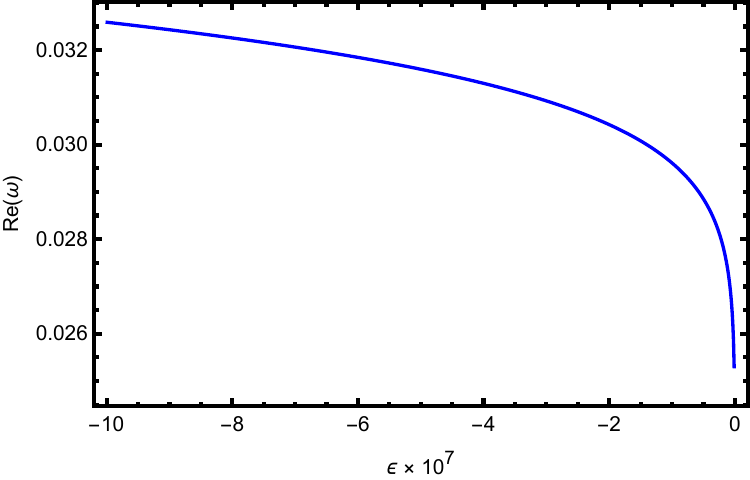}
\end{subfigure}
\begin{subfigure}{0.45\textwidth}
    \includegraphics[width=0.9\linewidth]{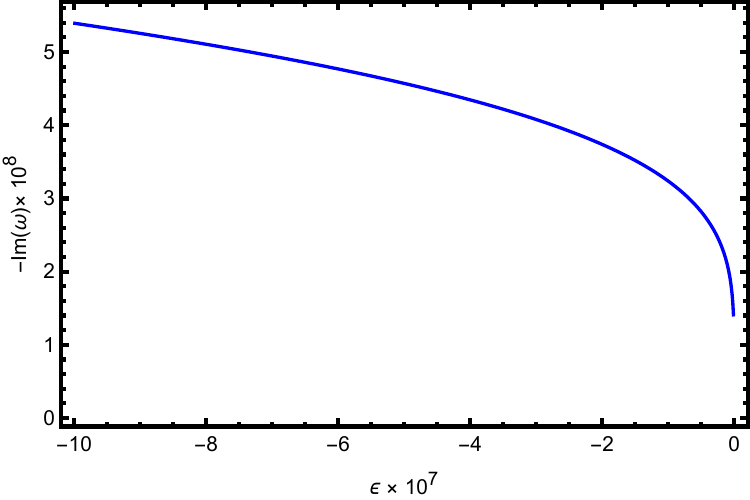}
\end{subfigure}
\caption{\textit{Left panel:} Plot of real part of the fundamental quasinormal frequency with $\epsilon$ for parametrized Damour-Solodukhin wormhole. Here $\epsilon$ is the parameter appearing in the parametrization of the redshift function \ref{A}. Now in case of Damour-Solodukhin wormhole we have $\lambda^{2}+1=\frac{1}{1+\epsilon}$. So this figure shows that how real part of quasinormal frequency varies with $\lambda^{2}$.~\textit{Right panel:}~Plot of imaginary part of the fundamental quasinormal frequency with $\epsilon$ for parametrized DS wormhole.}
\label{QNM-Isolated Wormhole}
\end{figure}

\begin{figure}[h!]
\begin{subfigure}{0.45\textwidth}
 \includegraphics[width=0.9\linewidth]{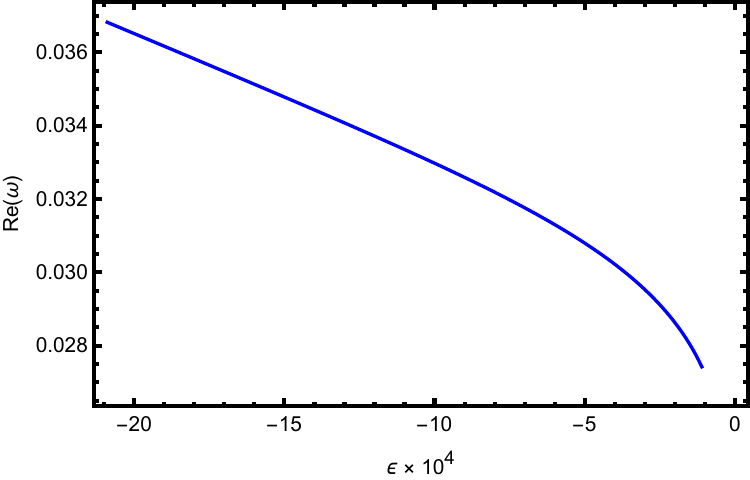}
\end{subfigure}
\begin{subfigure}{0.45\textwidth}
\includegraphics[width=0.9\linewidth]{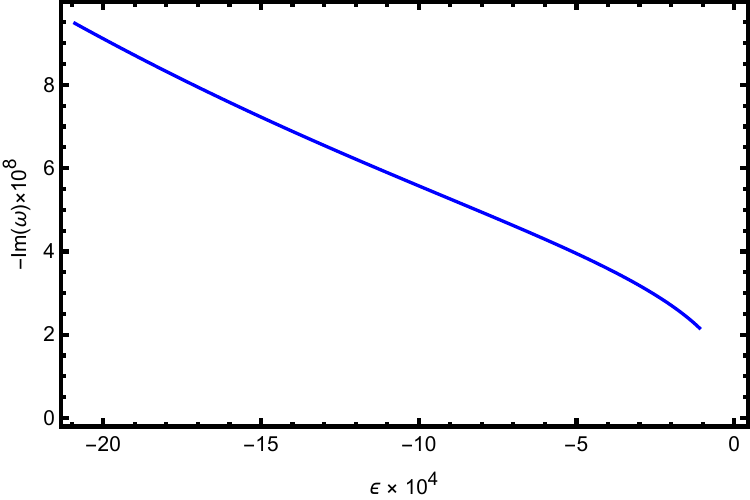}    
\end{subfigure}
\caption{\textit{Left panel:}~Plot of real part of the fundamental quasinormal frequency with $\epsilon$ for parametrized braneworld wormhole. Here $\epsilon$ is the parameter appearing in the parametrization of the redshift function~~\ref{A}. Now in our case of braneworld wormhole we have $\epsilon=-\frac{p}{(1+p)}$. So this figure shows that how real part of quasinormal frequency varies with $p$.~\textit{Right panel:}~Plot of imaginary part of the fundamental quasinormal frequency with $\epsilon$ for parametrized braneworld wormhole.}\label{fig: QNM Isolated brane}
\end{figure}

In this section, we will use the transfer matrix method \cite{Bueno:2017hyj} to calculate the quasinormal modes. In~~\ref{QNM-Isolated Wormhole},~ we have shown the variation of the real and imaginary parts of the quasinormal modes of the Damour-Solodukhin wormhole against $\epsilon$, which is the only non-zero BKP  parameter and related with $\lambda$. Since $\epsilon$ scales as $\lambda^{2}$, small values of $\epsilon$ provide even smaller values of $\lambda^{2}$. Since $L\simeq-4M~\rm Log[\frac{4}{\lambda^{2}}]$, a smaller value of 
$\lambda$ corresponds to a larger cavity length $L$ between the two potential barriers. Consequently, both the real and imaginary parts of the fundamental mode decrease \cite{Biswas:2022wah}, in agreement with the behavior shown in~~\ref{QNM-Isolated Wormhole}. An analogous trend is also found for the braneworld wormhole, as illustrated in~~\ref{fig: QNM Isolated brane}. We have also studied the dynamics of the perturbation in the time domain by Fourier transforming~~\ref{EM-master}. As an initial data, we use a Gaussian pulse centered at \(r_{s}=3M_{1}\) for the Damour-Solodukhin wormhole and at \(r_{s}=3M\) for the braneworld wormhole. In the two cases, the waveform is extracted numerically at \(r=25M_{1}\) and \(r=25M\), respectively. Denoting by \(\hat{\Psi}(r_{*},t)\) the Fourier-transformed master variable, the corresponding time-domain evolution equation takes the form
\begin{align}\label{master-time-domain}
 \frac{\partial^{2}\hat{\Psi}_{\ell m}}{\partial r_{*}^{2}}-\frac{\partial^{2}\hat{\Psi}_{\ell m}}{\partial t^{2}}+V_{\ell}(r)\hat{\Psi}_{\ell m}=0~.
\end{align}
The quasinormal-mode boundary conditions are imposed as \cite{Biswas:2022wah}
\(\partial_{t}\hat{\Psi}_{\ell m}=\pm\partial_{r_{*}}\hat{\Psi}_{\ell m}\) in the limits \(r_{*}\rightarrow\pm\infty\). To obtain the time-domain profile, we solve~~\ref{master-time-domain} with the initial conditions
\begin{align}\label{Time-Boundary-Condition}
 \hat{\Psi}_{\ell m}(t,r_{*}(r_{s}))=0~, \qquad
 \partial_{t}\hat{\Psi }_{\ell m}(t,r_{*}(r_{s}))=
 e^{-\frac{(r_{*}-r_{*}(r_{s}))^{2}}{\sigma^{2}}}~.
\end{align}

\begin{figure}[h!]
\begin{subfigure}{0.4\textwidth}
\includegraphics[width=1.0\linewidth]{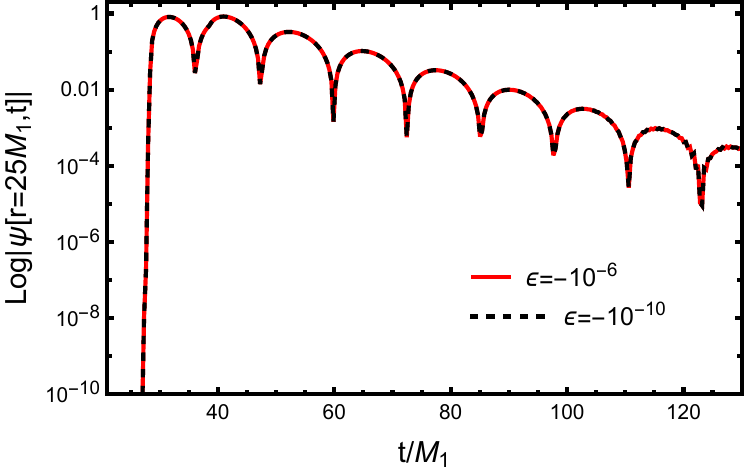}
\end{subfigure}
\begin{subfigure}{0.4\textwidth}
\includegraphics[width=1.0\linewidth]{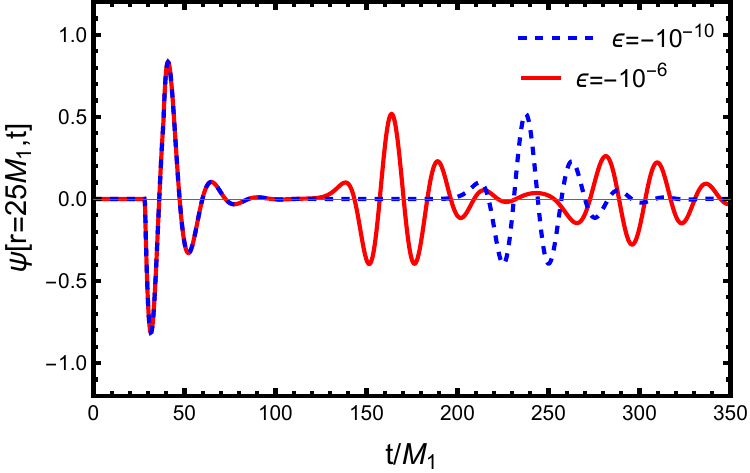}   
\end{subfigure}
\caption{Ringdown waveform of the parametrized Damour-Solodukhin wormhole under electromagnetic perturbation ($\ell=1$) for $\epsilon=10^{-10}$ and $\epsilon=10^{-6}$. Here we put the initial Gaussian impulse at $r=3M_{1}$ with $\sigma=10 M_{1}$, while the signal is observed at $r=25M_{1}$. \textit{Left Panel:} shows the log plot of the primary signal. It is mostly captured by the photon sphere modes, explaining less sensitivity towards variation of $\epsilon$. \textit{Right Panel:} shows that for small values $\epsilon$, the echo time delay increases, which is expected from \cite{Bueno:2017hyj}.}\label{DS-Ringdown}
\end{figure}

\ref{DS-Ringdown} displays the ringdown waveform of the parametrized Damour-Solodukhin wormhole under electromagnetic perturbations for \(\ell=1\). In the left panel, we plot the logarithm of the absolute value of the wavefunction and find that the primary signal is essentially insensitive to variations in \(\epsilon\). This behavior is expected, since the primary part of the ringdown is mainly governed by perturbations near the photon sphere \cite{Cardoso:2016rao}, which in the present case is independent of \(M_{2}\), and hence also independent of \(\epsilon\). By contrast, the echo time delay does respond to changes in the throat length \(L\), which is controlled here by \(\epsilon\). As expected \cite{Bueno:2017hyj}, the echo time delay increases as \(\epsilon\) decreases. This is clearly visible in the right panel of ~~\ref{DS-Ringdown}.

\begin{figure}[h!]
\begin{subfigure}{0.4\textwidth}
\includegraphics[width=1.0\linewidth]{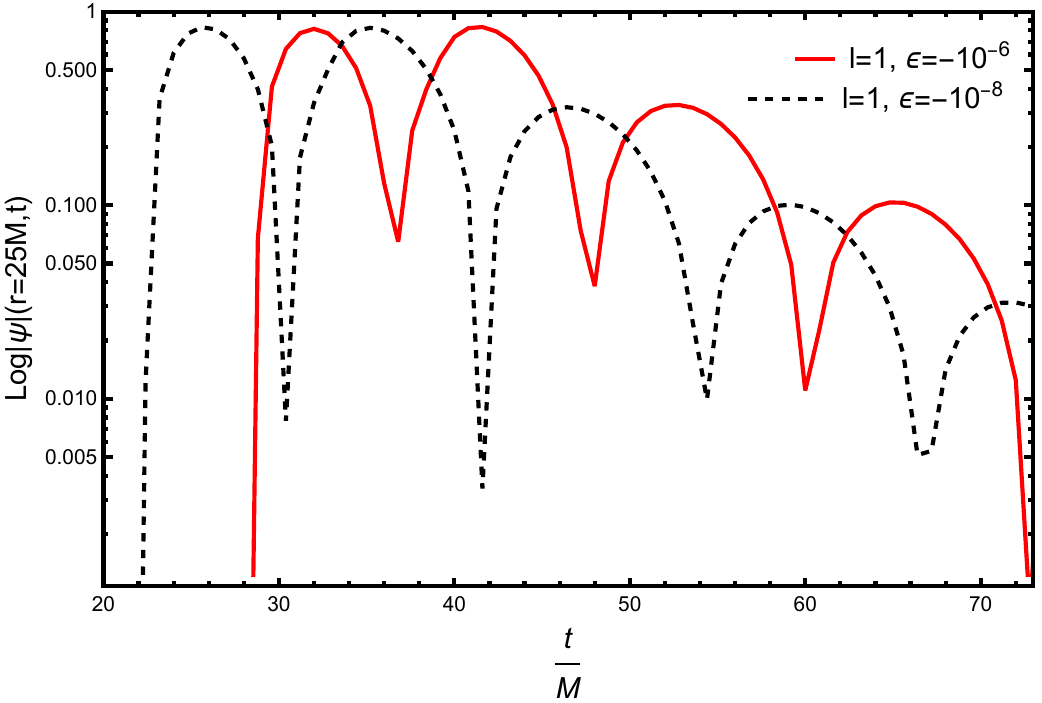}
\end{subfigure}
\begin{subfigure}{0.4\textwidth}
\includegraphics[width=1.0\linewidth]{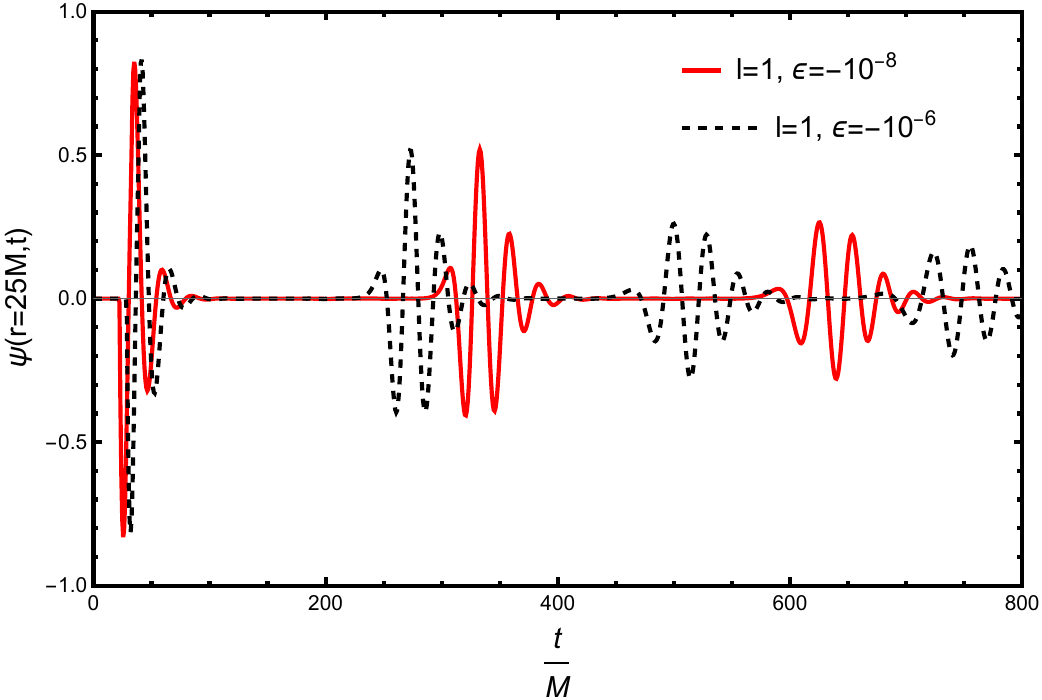}   
\end{subfigure}
\caption{Ringdown waveform of parametrized braneworld wormhole under electromagnetic perturbation ($\ell=1$) for $\epsilon=10^{-8}$ and $\epsilon=10^{-6}$. We put the initial Gaussian impulse at $r=3M$ with $\sigma=10 M$. While the signal is observed at $r=25M$. \textit{Left Panel:} Shows the log plot of the primary signal. \textit{Right Panel:} Shows the entire ringdown waveform which includes subsequent echoes after the primary signal.} \label{Ringdown-Brane}
\end{figure}

The corresponding ringdown waveform for the parametrized braneworld wormhole is shown in~~\ref{Ringdown-Brane}. In this case, unlike the Damour-Solodukhin wormhole, even the primary signal depends on \(\epsilon\), as seen from the left panel. The variation in the echo time delay is again evident from the right panel. There are two possible reasons for this behavior. First, the metric component \(g_{tt}\) of the braneworld spacetime depends explicitly on \(p\), and therefore on \(\epsilon\). Second, the BKP parametrization may not adequately capture the near-throat region where the photon sphere is located. The second possibility is particularly interesting. From ~\ref{braneworld-metric-comparison}, one finds that the parametrized \(g_{tt}\) agrees well with the exact metric for both values of \(\epsilon\). This raises the possibility that the primary signal is sensitive to the parametrization itself. To examine this point further, in~~\ref{sec-8}, we compute the primary waveform of the galactic wormhole while neglecting one near-field parameter, \(a_{1}\); see ~\ref{fig:without a1}. 

\section{Parameterized galactic Damour-Solodukhin Wormhole}\label{sec-5}
So far, we have discussed the parametrized wormholes without an environment. Since a substantial fraction of a galaxy’s mass budget is attributed to dark matter, it is physically motivated to examine wormholes embedded in a galactic halo. In particular, in continuation of our previous study, we will be discussing the galactic Damour-Solodukhin WH as proposed in \cite{Biswas:2023ofz}. The authors of \cite{Biswas:2023ofz} have demonstrated the wormhole inside the galactic halo in a fully relativistic setting where the dark matter halo is described by an anisotropic fluid with vanishing radial pressure whose density follows Hernquist-type profile~\cite{hernquist1990analytical},
\begin{align}
    \rho(r)=\frac{Ma}{2\pi r(r+a)^{3}}~,
\end{align}
where $M$ is the mass of the galactic halo, and $a$ is the typical size of the halo. It was shown that the Damour-Solodukhin wormhole inside the halo is described by the line element \ref{General-static-metric}, with the following metric components,
\begin{align}\label{ga-DS}
    &h(r)=1-\frac{2}{r}\left[M_{2} +\frac{Mr^{2}}{(r+a)^{2}}\left(1-\frac{2M_{2}}{r} \right) \left(1-\frac{2M_{1}}{r} \right)\right] \quad \text{with}\\
   & f(r)=\left(1-\frac{2M_{1}}{r} \right)e^{\gamma}    
\end{align}
where $\gamma=\sqrt{\frac{M}{2a+4M_{1}-M}}\{2~\tan^{-1}\left( \frac{a+r-M}{\sqrt{M(2a-M+4M_{1})}}\right)-\pi \}$. From the above metric elements, it is clear that for large $r$, the mass profile is dominated by that of the Hernquist profile, and only near the throat, $r=2M_{2}$, it is dominated by $M_{2}$. Below we will be describing the parametrization of this wormhole with the scheme introduced in~~\ref{sec2}.\\

\begin{figure}[h!]
    \begin{subfigure}{0.45\textwidth}
        \includegraphics[width=0.9\linewidth]{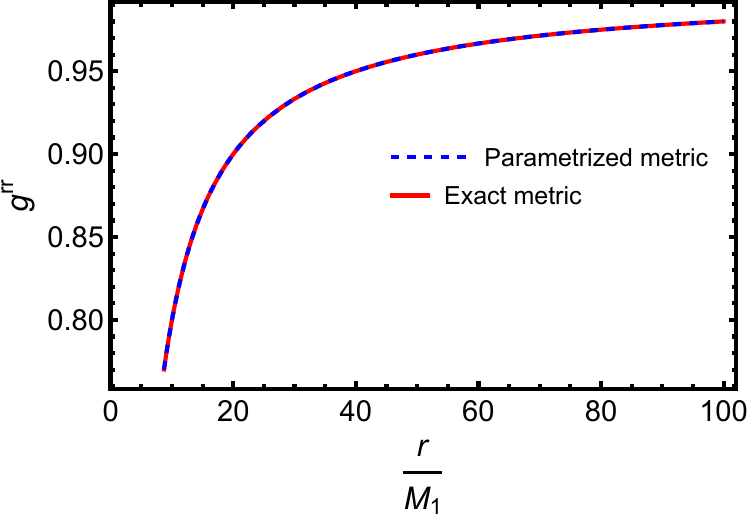}
    \end{subfigure}
    \begin{subfigure}{0.45\textwidth}
        \includegraphics[width=0.9\linewidth]{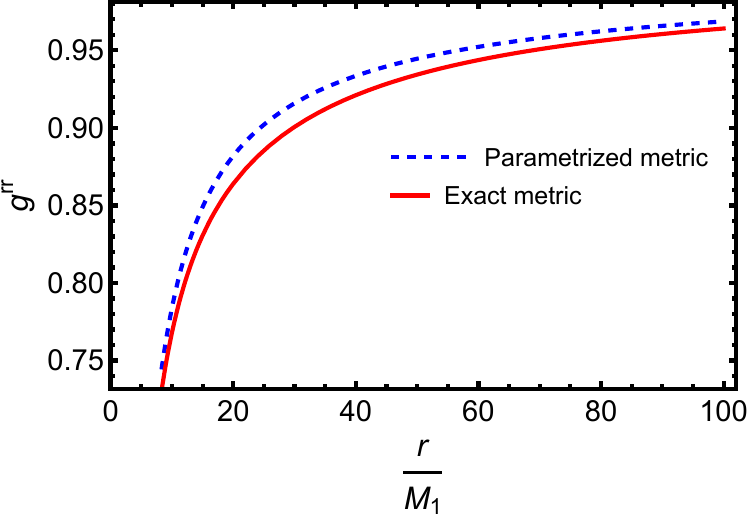}
    \end{subfigure}
    \caption{Comparison of $g^{rr}$ of exact spacetime with that of parametrized metric for two sets of parameters.~\textit{Left panel}: In this figure we have compared $g^{rr}$ of exact spacetime with that of parametrized metric for the parammeters $a=10^{8}M_{1}$, $M=10^{4}M_{1}$.~\textit{Right panel}: In this figure we have compared $g^{rr}$ of exact spacetime with that of parametrized metric for the parammeters $a=10M_{1}$, $M=M_{1}$. From this it is clear that as $\frac{M}{a}$ ratio decreases the parametrization overlaps with that of the exact metric component.}\label{fig:grr-comp}
\end{figure}

\indent Our galactic wormhole is described by the following four parameters: $M_{1}$, $\lambda^{2}$, $\frac{M}{M_{1}}$, and $\frac{a}{M_{1}}$. 
Due to the algebraic nature of $h(r)$ it can be easily parametrized and the first four BKP parameters are given by,
\begin{align}
    & h_{0}=0~\label{h-gal-ds-1},\\
    & b_{0}=\frac{M}{M_{2}}~\label{h-gal-ds-2},\\
  & b_{1}=\frac{a^2 M+4 a M M_{2}+4 M M_{1} M_{2}}{M_{2} (a+2 M_{2})^2}~\label{h-gal-ds-3},\\
  & b_{2}=-\frac{a(a^2+6 a M_{2}+8 M_{1} M_{2})}{(a+2 M_{2}) \left(a^2+4 a M_{2}+4 M_{1} M_{2}\right)}~\label{h-gal-ds-4}.
\end{align}
One can verify that these parameters are consistent with the isolated case (that is, without dark matter environment), that is, all of them vanish for $M=a=0$. From the above expressions, it is clear that, unlike the isolated Damour-Solodukhin wormhole, this time we cannot fix all the BKP parameters just by specifying one of the galactic parameters namely, $\lambda^{2}$, $\frac{M}{M_{1}}$, and $\frac{a}{M_{1}}$. However, upon defining $q=\frac{M}{M_{1}}$ and $y=\frac{M}{a}$ and assuming that  $\lambda^{2}\ll 1 $, one finds that the following inequalities must be satisfied by the parameters if those correspond to the galactic Damour-Solodukhin wormhole:   
\begin{align}
 b_{0}<q~,\label{b0-b1-constraint-1}\\
 b_{1}< q-\frac{4q \lambda^{2}}{\left(\frac{q}{y}+2\right)^{2}}~\label{b0-b1-constraint-2}.
\end{align}
A similar inequality can likewise be derived for \(b_{2}\). These inequalities should be regarded as theory-dependent inputs, since the right-hand side is determined entirely by the specific model under consideration. In~~\ref{fig:grr-comp}, we compare the parametrized metric with the exact spacetime and find that the agreement improves as the ratio \(M/a\) decreases. In particular, the parametrization reproduces the exact behavior more accurately at large distances. This is expected, since the far-field sector has been incorporated through the relevant asymptotic parameter, namely, \(b_{0}\).

The situation is different in the near-throat region. In principle, this region is controlled by an infinite hierarchy of near-field parameters \(b_{i}\), whereas in the present analysis, we retain only \(b_{1}\) and \(b_{2}\). The comparison in ~~\ref{fig:grr-comp} shows that, as \(M/a\) increases, the effect of dark matter modifies the near-throat geometry relative to the isolated Damour-Solodukhin wormhole more significantly. In that regime, the truncated parametrization begins to deviate from the exact spacetime, indicating that additional near-field coefficients are required for an accurate match. By contrast, when \(M/a \ll 1\), the parametrized metric component \(g_{rr}\) remains in good agreement with the exact solution.

\begin{figure}[h!]
    \begin{subfigure}{0.45\textwidth}
        \includegraphics[width=0.9\linewidth]{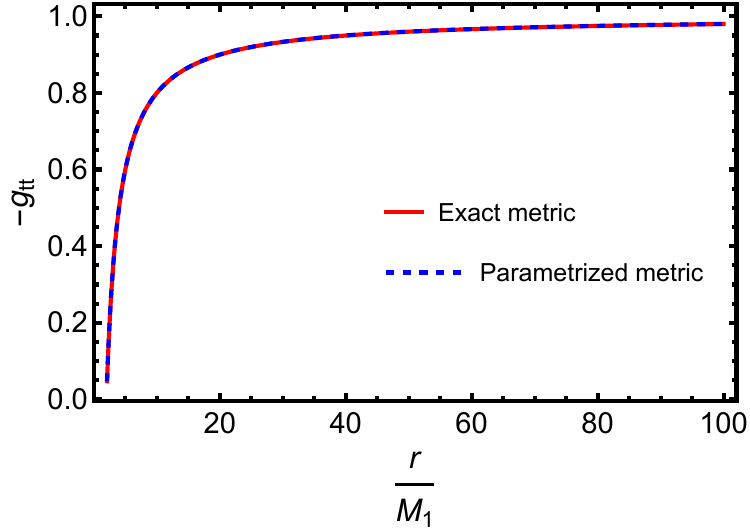}
    \end{subfigure}
    \begin{subfigure}{0.45\textwidth}
        \includegraphics[width=0.9\linewidth]{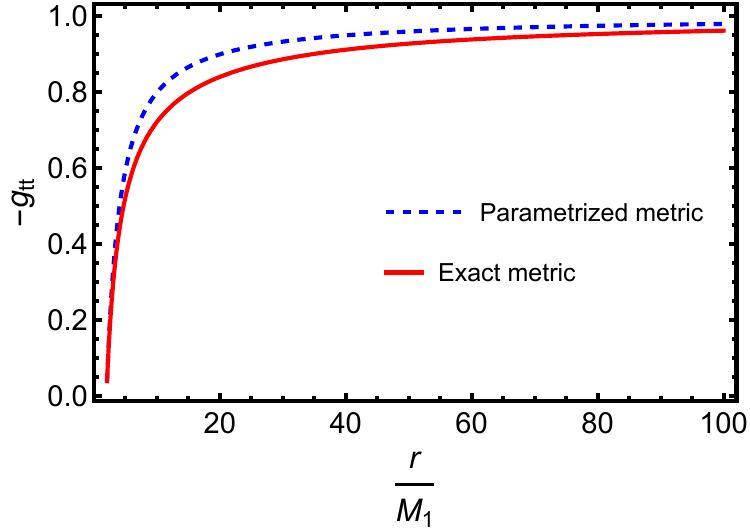}
    \end{subfigure}
    \caption{Comparison of $g_{tt}$ of exact spacetime with that of parametrized metric for two sets of parameters.~\textit{Left panel}: In this figure we have compared $g_{tt}$ of exact spacetime with that of parametrized metric for the parammeters $a=10^{8}M_{1}$, $M=10^{4}M_{1}$.~\textit{Right panel}: In this figure we have compared $g_{tt}$ of exact spacetime with that of parametrized metric for the parammeters $a=10M_{1}$, $M=M_{1}$. From this it is clear that as $\frac{M}{a}$ ratio decreases, the parametrization overlaps with that of the exact metric component.}\label{fig:gtt-comp}
\end{figure}

In what follows, in order to keep the computation tractable, we restrict ourselves to the observationally viable regime \(M/a \ll 1\) and retain only these two near-field parameters. This completes the parametrization of \(g_{rr}\).
Although the transcendental form of $g_{tt}$ does not obstruct coefficient extraction in principle, the direct BKP reconstruction leads to coefficients that violate the required hierarchy or produce unphysical behavior in the astrophysical regime. In what follows, we will use a controlled approximation valid for $a \gg M \gg   M_1$ (see Appendix A for details on the issues). To circumvent this pathology, we approximately parametrize the spacetime as follows. We observe that $A_{iso}(x\rightarrow 1 )=A_{gal}(x\rightarrow 1 )= 1$. This happens because $\gamma(r\rightarrow \infty)=0$, so that in the asymptotic region we can approximate the $g_{tt}$ of the galactic wormhole with that of the isolated one. Using this trick, we can parametrize the far region of the galactic wormhole, and it yields,
\begin{align}\label{Gal-f-new-trick}
  &\epsilon=-\frac{\lambda^{2}}{1+\lambda^{2}}~,\\
 & a_{0}=0.
\end{align}
Next, to parametrize the near throat region of the wormhole, we observe that for typical sets of galactic parameters $\gamma\ll 1$, more precisely, this  happens when the hierarchy $a\gg M\gg M_{1}>0$ holds. In that case, we approximate $e^{\gamma}\simeq 1+\gamma$ ~~\footnote{The reader may argue that $\gamma$ itself contains transcendental behavior, but since expansion of $tan^{-1}(x)$ has polynomial dependence on $x$ so that one can read-off $a_{1}$.}~~and determine the near throat parameter $a_{1}$ as,
\begin{align}\label{a1-new-trick}
    a_{1}=\frac{N}{D}~,
\end{align}
where, \begin{align}
    &N=-2 \alpha  (2 M_{1}-3 M_{2}) \left(a^2+4 a M_{2}+4 \left(M (M_{1}-M_{2})+M_{2}^2\right)\right) \sigma+\pi  \alpha  a^2 (2 M_{1}-3 M_{2})+4 a M_{2} (\pi  \alpha  (2 M_{1}-3 M_{2})\\
    \nonumber
    &-2 M (M_{1}-M_{2}))+4 \left(M^2 M_{2} (M_{1}-M_{2})+M (M_{1}-M_{2}) (2 \pi  \alpha  M_{1}-4 M_{1} M_{2}-3 \pi  \alpha  M_{2})+\pi  \alpha  M_{2}^2 (2 M_{1}-3 M_{2})\right)
\end{align}
here $\alpha=\sqrt{M (2 a-M+4 M_{1})}$, $\sigma=\tan^{-1}\left(\frac{a+2M_{2}-M}{\alpha}\right)$ and $D= M_{2} (2 a - M + 4 M_{1})(a^{2} + 4 a M_{2} + 4 (M (M_{1} - M_{2}) + M_{2}^{2}))$. We calculate $f_{0}$ from it's definition, that is,
\begin{align}
    f_{0}=f(r=2M_{2})~.
\end{align}
In principle, one can calculate other near field parameters, but to avoid computational expense, we will include only $a_{1}$. The parameter $a_{1}$ also has correct correspondence with the isolated case, as it vanishes when $M\rightarrow 0$. In \ref{fig:gtt-comp} we have compared our parametrization with that of the exact metric. As in the case of $g_{rr}$, the parametrization of $g_{tt}$ works well when $\frac{M}{a}\ll 1$.

\section{Consistency with Shadow Observation}\label{sec-6}
Before we proceed to the computation of quasinormal modes and ringdown waveform for the parametrized galactic Damour-Solodukhin wormhole, in this section, we will verify whether our working parameter space for those are consistent with the shadow observations, which solely depend on the $g_{tt}$ component of the metric. For this purpose, we first note that the shadow radius for the galactic Damour-Solodukhin wormhole is given by \cite{Biswas:2023ofz},
\begin{align}\label{DS-Shadow}
    r_{\rm sh}=3\sqrt{3}M_{1}\left[1+\frac{M}{a}+\frac{M(5M-18M_{1})}{6a^{2}}\right]~.
\end{align}
If we assume that $M\gg M_{1}$,~then by defining galactic compactness, $C_{g}=\frac{M}{a}$ we get,
\begin{align}\label{shadow-compactness}
    \frac{r_{\rm sh}}{3\sqrt{3}M_{1}}\simeq\left[1+C_{g}+\frac{5C_{g}^{2}}{6}\right]~.
\end{align}
Now demanding that the above dimensionless quantity satisfies the upperbound of shadow at $1\sigma$ level as given in \cite{Vagnozzi:2022moj}, we get,
\begin{align}\label{gal-compactness-by-shadow}
   0< C_{g} < 0.005~.
\end{align}
On the other hand, from the accretion consideration of the rotating Damour-Solodukhin wormhole\cite{Karimov:2019qfw}, it is known that the bound on $\lambda $ is $\lambda < 10^{-3}$. Therefore, consistency with shadow observations, which will give a valid parameter space for the BKP parameters, proceeds as follows: pick up $M$ and $a$, then check if \ref{gal-compactness-by-shadow} is satisfied. Then, with an appropriate value of $\lambda$, one can compute the values of BKP parameters $(b_{0},b_{1},b_{2})$ by using the formulas \ref{h-gal-ds-1},\ref{h-gal-ds-2},\ref{h-gal-ds-3} and \ref{h-gal-ds-4}. If the parametrization is correct, then \ref{b0-b1-constraint-1} and \ref{b0-b1-constraint-2} will be satisfied. Therefore, if from the shadow observations (which only depend on $g_{tt}$ and it's derivative) one can determine a valid set of galactic parameters, then using the above procedure one can determine valid ranges of $(b_{0},b_{1},b_{2})$, which will provide a parametrization of $g_{rr}$. 
One can also proceed in another way by directly calculating the formula for the outer light ring radius using the parametrized metric through~\ref{A} by exploiting the condition $r_{\rm ph}f^{\prime}(r_{\rm ph})=2f(\rm r_{ph})$~. Then the largest root of this equation can be used to compute the shadow radius as
\begin{align}\label{shadow}
    \frac{1}{r^{2}_{\rm sh}}=\frac{f(r_{\rm ph})}{r_{\rm ph}^{2}}~.
\end{align}

This equation shows that the shadow radius will be a complicated function of $\epsilon$, $ a_{0} $, $ a_{1} $ and $f_{0}$.
Therefore, from the shadow observations, one can determine the BKP parameters $\epsilon$, $ a_{0} $, $ a_{1} $ and $f_{0}$. With these and using the relations of these parameters with the galactic parameters, one can numerically find the range of galactic parameters $(M,a)$ for which \ref{gal-compactness-by-shadow} is satisfied. Next, upon using this range of $(M,a)$ and then demanding consistency with accretion physics (which gives a bound on $\lambda$), one can determine the parameters associated with $g_{rr}$. 

This completes an observationally valid parametrization of the galactic Damour-Solodukhin wormhole. In this paper, we will consider the following parameters for a galaxy which satisfies \ref{gal-compactness-by-shadow}: 1) $a=10^{8}M_{1},M=10^{4}M_{1}$, 2) $a=10^{8}M_{1},M=10^{2}M_{1}$. Interestingly, these parameters satisfy \ref{b0-b1-constraint-1} and \ref{b0-b1-constraint-2} when $\lambda< 10^{-3}$.
\begin{figure}[htbp!]
    \centering
    \includegraphics[width=0.44\textwidth]{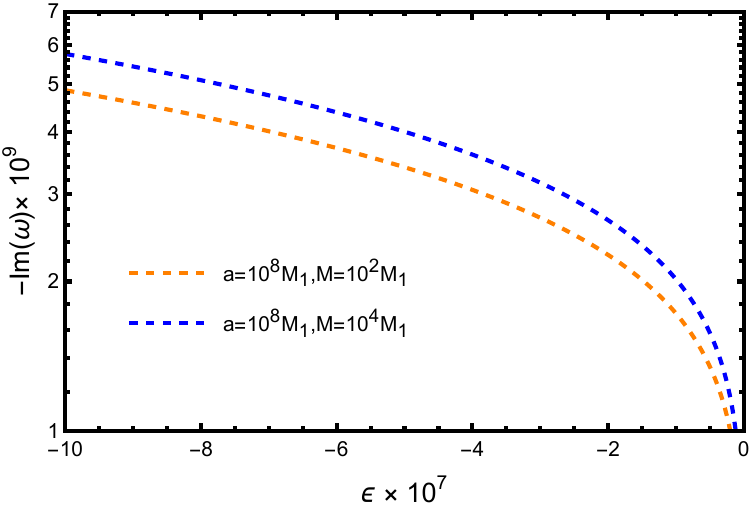}
    \includegraphics[width=0.45\textwidth]{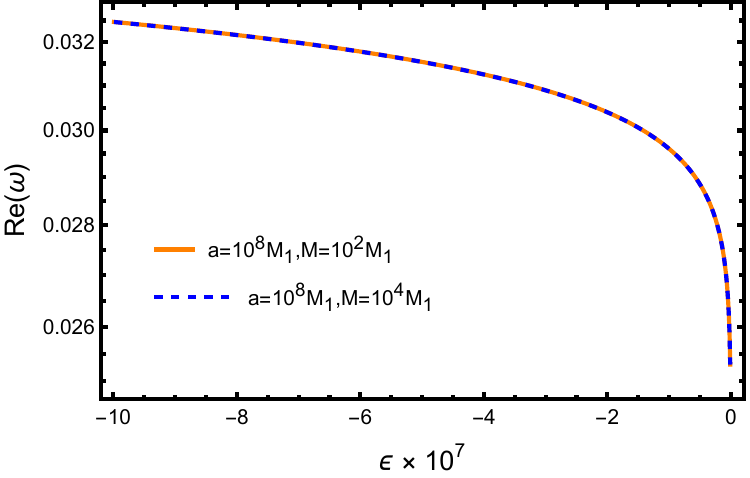}
    \caption{Plot of real (left panel) and imaginary (right panel) part of fundamental quasi normal frequencies with $\epsilon$. Here we have plotted quasi normal frequencies for two sets of galactic compactness, namely $C_{g}=10^{-4}$ and $C_{g}=10^{-6}$. From right figure we can infer that as the galactic compactness increases the galactic wormhole becomes more stable. }
    \label{fig:QNM-GAL}
\end{figure}

\section{Quasinormal Ringing of the parametrized wormhole}\label{sec-7}
\begin{figure}
    \centering
    \includegraphics[width=0.5\linewidth]{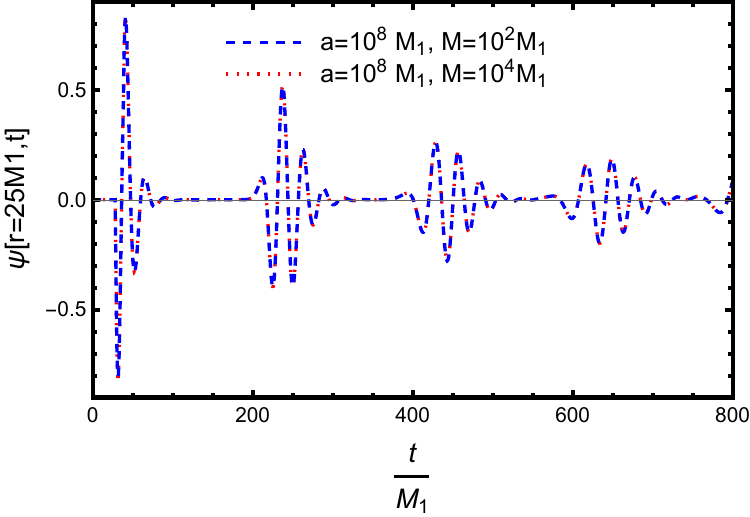}
    \caption{Ringdown waveform of parametrized galactic Damour-Solodukhin wormhole for two sets of galactic parameter with $\epsilon=-10^{-10}$. Here we set the angular momentum index $\ell=1$ and we put the initial Gaussian impulse at $r=3M_{1}$ while the observation is done at $r=25M_{1}$.}
    \label{fig:ringdown-galds}
\end{figure}

\begin{figure}[h!]
    \begin{subfigure}{0.45\textwidth}
        \includegraphics[width=0.9\linewidth]{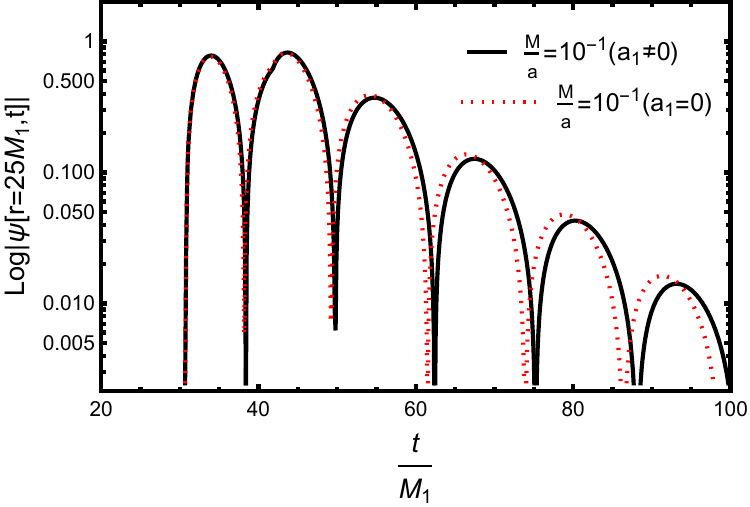}
    \end{subfigure}
    \begin{subfigure}{0.45\textwidth}
        \includegraphics[width=0.9\linewidth]{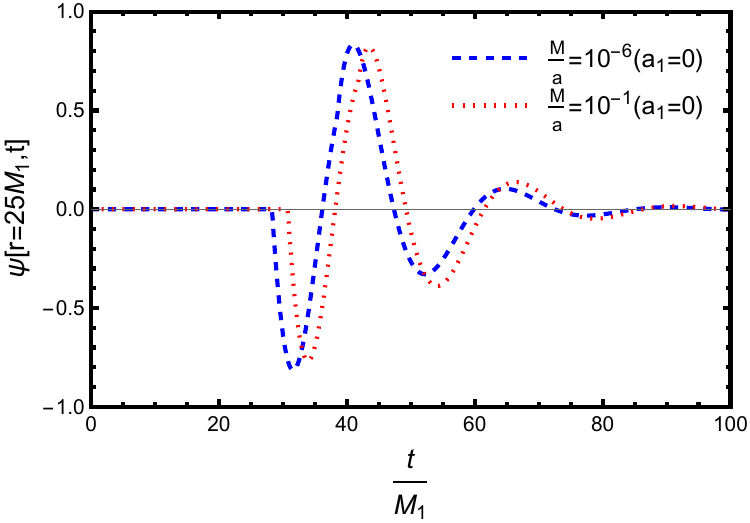}
    \end{subfigure}
    \caption{This figure compares the primary signal for the galactic wormhole with that of the fictitious metric obtained by setting $a_{1}=0$.~\textit{Left panel:} Shows the comparison in the log-scale for the galactic compactness $\frac{M}{a}=\frac{1}{10}$. \textit{Right panel:} Shows how the primary signal changes even for the fictitious metric for various galactic compactness.}\label{fig:without a1}
\end{figure}

Having discussed the consistency of the working parameter space, we are now in a position to study the quasinormal ringing of the parametrized wormhole. Following the procedure given in \cite{Bueno:2017hyj,Biswas:2023ofz}, we have used the transfer matrix method to compute the fundamental modes for various galactic parameters and $\lambda$. In~\ref{fig:QNM-GAL} we have plotted the real and the imaginary part of the fundamental quasinormal modes for various galactic parameters against $\epsilon$. From the left panel of this figure, it is clear that imaginary part is sensitive to galactic compactness, in particular, as the galactic compactness increases, the imaginary part increases and therefore in the presence of the dark matter the wormhole becomes more stable under linear perturbation. This also confirms the previous results in\cite{Biswas:2023ofz}. However, the change in the imaginary part is very small and goes as $M_{1}|\text{Im}(\Delta\omega)|\sim\mathcal{O}(10^{-9})$. To obtain the ringdown waveform, we have followed the procedure illustrated in \ref{sec-4-2} of \ref{sec-4-0}.~\ref{fig:ringdown-galds} shows the ringdown waveform of the parametrized galactic Damour-Solodukhin wormhole. At the end of \ref{sec-4-0}, we commented that the inability to parametrize the near throat region of the braneworld wormhole might have caused the variation of the primary signal with $\epsilon$. To see this issue more prominently, in \ref{fig:without a1}, we have compared the primary signal of the galactic wormhole with that of a fictitious metric which is obtained by setting the leading near-field parameter $a_{1}=0$. In the left panel, we have compared these two primary signals in the log scale and in the right panel, we have compared how the signal is modified even for the fictitious metric for two different galactic compactnesses. This shows that the absence of the leading near-field parameter can indeed modify the primary signal, which may be detectable. It should be noted that, the value $M/a=10^{-1}$
 used in \ref{fig:without a1} is not part of the shadow-allowed parameter space; it is introduced only as a diagnostic case to make the effect of removing the leading near-field coefficient visually apparent. In the next section, we will explore more on this issue.
\section{Sensitivity of the primary signal to the parametrization}\label{sec-8}
\begin{figure}[h!]
    \begin{subfigure}{0.6\textwidth}
    \includegraphics[width=0.9\linewidth]{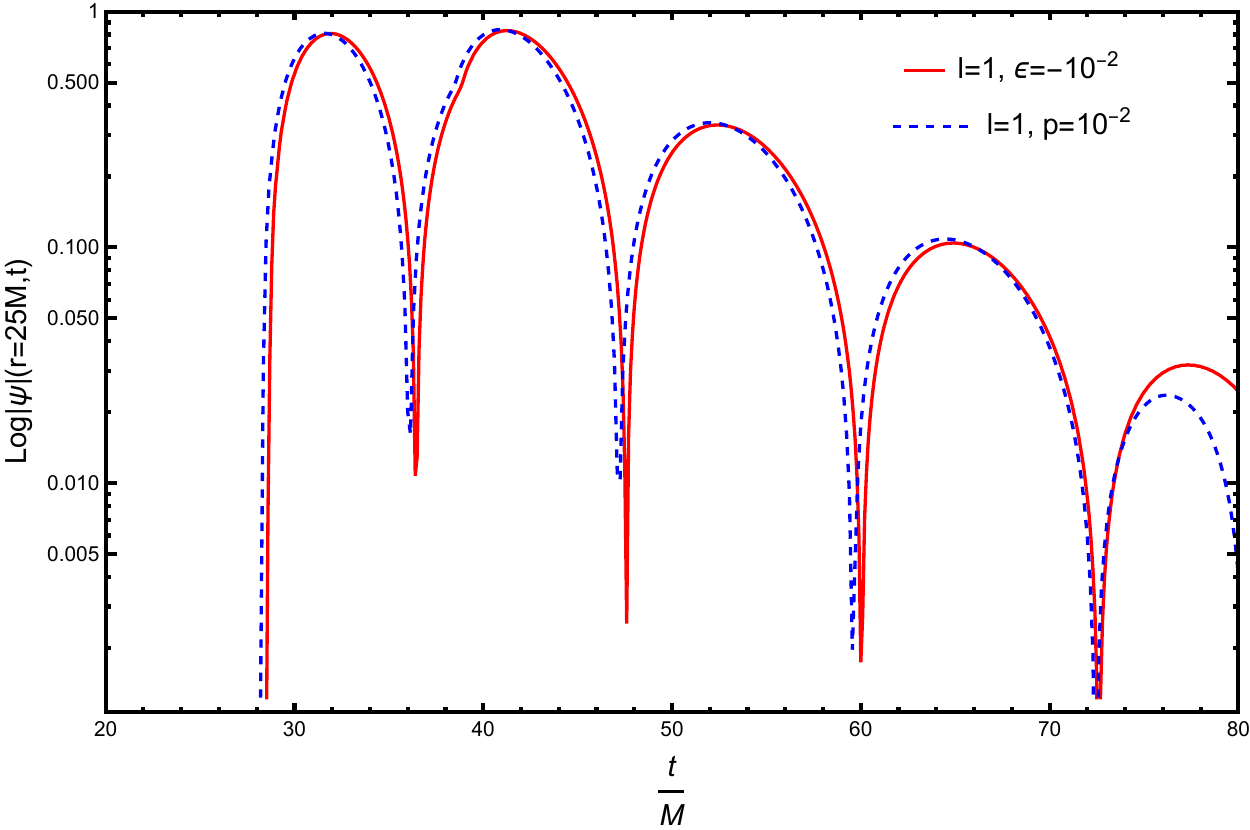}
    \end{subfigure}
    \begin{subfigure}{0.6\textwidth}
    \includegraphics[width=0.9\linewidth]{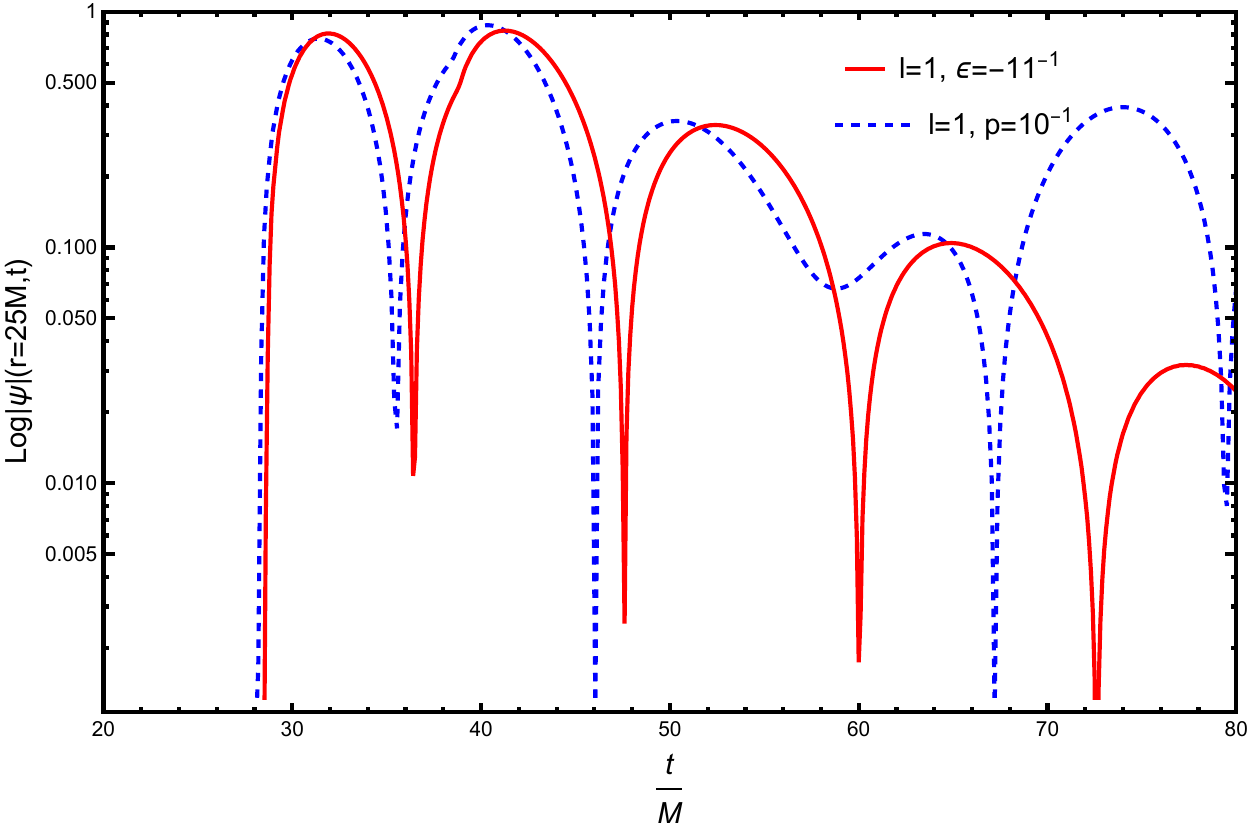}
    \end{subfigure}
    \caption{In this figure we have analysed the contamination of the primary signal of the braneworld wormhole for $\ell=1$ under parametrization. Here the \textit{blue} dashed curve denotes the primary signal corresponding to the exact spacetime and the \textit{red} curve denotes the parametrized one.~In the comparison, we have used the one-to-one correspondence of $p$ and $\epsilon$ (see~~\ref{brane-param}) and the perturbing potential~~\ref{EM-pot}.~\textit{Top panel}:~Here we made the comparison for $p=10^{-2}$~(\textit{for exact case}) or equivalently $\epsilon\approx-10^{-2}$~(\textit{for parametrized case}).~\textit{Bottom panel}:~Here we made the comparison for $p=10^{-1}$~(\textit{for exact case}) or equivalently $\epsilon=-11^{-1}$~(\textit{for parametrized case}). This comparison shows that as $p$ decreases the exact signal matches well with that of the parametrized signal.}\label{brane-primary-signal-comparison}
\end{figure}

As discussed near the end of~~\ref{sec-4-0}~~and in \ref{sec-7},the primary part of the ringdown waveform is sensitive to the parametrization in the sense that if we do not include near field parameters (such as $a_{1}$), then the primary signal differs slightly. In this section, we will investigate this issue further. For that purpose, we recall from \cite{Cardoso:2016rao,Cardoso:2008bp,Rosato:2025lxb}, that the primary part of the ringdown signal can be mimicked by the so called \textit{photon sphere (light ring)} modes, namely
\begin{align}\label{Light-ring-modes}
    \omega_{\rm L}=\Omega_{c}\ell-i\left(n+\frac{1}{2}\right)|\Lambda_{L}|~\quad n=0,1,2,...,
\end{align}
where $\Omega_{c}$ is the angular velocity of the photons at the circular null geodesics, and $\Lambda_{L}$ is the Lyapunov time scale. For a static spherically symmetric spacetime, these quantities are given by,
\begin{align}
    &\Omega_{c}=\sqrt{\frac{f(r_{\rm ph})}{r_{\rm ph}^{2}}}~\quad \text{and}\\
    &\Lambda_{L}=\frac{1}{\sqrt{2}}\sqrt{-\frac{r_{\rm ph}^{2}}{f(r_{\rm ph})}\left(\frac{d^2}{dr_*^2}\frac{f}{r^2}\right)_{r=r_{\rm ph}}}~.
\end{align}
Here, the radius $r_{\rm ph}$ is given by solving the equation $r_{\rm ph}f^{\prime}(r_{\rm ph})=2f(r_{\rm ph})$~. The above formula \ref{Light-ring-modes} is one of the main motivations behind expecting that horizon-less compact objects~\cite{Cardoso_2019} can indeed show similar ringdown as that of black holes. This result can be obtained by considering the eikonal limit ($\ell \gg 1$) of the standard WKB formula~\cite{schutz1985black} to obtain the QNMs of black holes. Although \ref{Light-ring-modes} is derived in the eikonal limit $\ell\gg 1$, it can also provide useful estimates for smaller values of $\ell$ \cite{Berti:2005eb,Iyer-QNM}. To this end, we emphasize that \textit{$\omega_{\rm L}$'s are not quasinormal modes of the wormhole spacetime but, as shown in \cite{Rosato:2025lxb}, that these modes (in the reference, it is denoted as $\omega^{\rm BH}$) can be used to reconstruct the primary part of the ringdown signal}. In what follows, we will compute $\omega_{\rm L}$'s considering $n=0$ (\textit{fundamental mode}) for the case of exact and parametrized spacetime and compare them to see how the parametrization can affect the light ring modes.\\
\\
\indent In~~\ref{comp-light-right-location}, we have considered the light ring location~($r_{\rm ph}$) and redshift factor~($f(r)$)~ at the location of light ring for the case of braneworld wormhole. We have compared these quantities with those of the exact spacetime.  While making such comparison, we pick a value of $p$ and then obtained the corresponding parameters (only far field) using \ref{brane-param}. With these, we have computed the light ring location~($r_{\rm ph}$) and redshift factor~($f(r)$) both for the case of exact and parametrized spacetime. The table shows that as $p$ decreases these quantities for the parametrized spacetime start matching with those of the exact spacetime. This is also in agreement with our previous conclusion from \ref{braneworld-metric-comparison}. We have further verified these conclusions from the perspective of linear perturbation. For that purpose, we have computed $\omega_{\rm L}$ and the primary waveform for both the cases of exact and parametrized spacetime. In \ref{omwga-L-brane}, we have compared $\omega_{\rm L}$'s and in \ref{brane-primary-signal-comparison}, we do the same for the primary signal. All these lead to the conclusion:~\textit{inability to parametrize the near throat region of the wormhole will generally affect the primary signal.}\\
\indent We have further extended our analysis toward the case of a galactic Damour-Solodukhin wormhole. However, as mentioned in \ref{sec-5}, this time there is no one-to-one correspondence between parameters of the spacetime (that is $\lambda^{2}$, $\frac{M}{M_{1}}$ and $\frac{a}{M_{1}}$) with the BKP parameters. This leads us to adopt a slightly different methodology for comparison. Here, we have first fixed $\lambda=10^{-3}$, and then for various values of $(M,a)$,  obtained the BKP parameters. With these in \ref{omega-L-Gal}, we have compared $\omega_{\rm L}$ for the exact and parametrized spacetimes. This shows that as the galactic compactness, $\frac{M}{a}$ decreases, the frequencies match quite well. Moreover, in \ref{sec-6} we have made an analogous situation of the braneworld case by  setting $a_{1}=0$ by hand. Here we have computed $\omega_{\rm L}$ for such fictitious spacetime also. As in this case the $\omega_{\rm L}$s' are different from the parametrized wormhole, so that it gives us a justification for \ref{fig:without a1}. The absolute relative errors in~\ref{comp-light-right-location},
\ref{omwga-L-brane}, and \ref{omega-L-Gal} quantify the
fidelity of the parametrized description. For the braneworld case, the mismatch
in $r_{\rm ph}/M$, $f(r_{\rm ph})$, and $\omega_{\rm L}$ decreases rapidly as
$p$ is reduced, while for the galactic Damour-Solodukhin wormhole the
light-ring-mode errors are already below $10^{-2}\%$ for
$M/a\leq10^{-4}$ and decrease further with galactic compactness. The comparison
with the fictitious $a_1=0$ case also shows that the primary/light-ring response
retains measurable sensitivity to the leading near-throat coefficient.

\begin{table}[t] \centering \begin{tabular}{|c|c| c| c |c| c| c|} 
\hline $p$ & $r_{\rm ph}^{\rm para}/M$ & $f(r_{\rm ph}^{\rm para})$ & $r_{\rm ph}^{\rm exact}/M$ & $f(r_{\rm ph}^{\rm exact})$ & $\delta_r(\%)$ & $\delta_f(\%)$ \\ \hline $10^{-1}$ & $2.97426$ & $0.324099$ & $2.84496$ & $0.343799$ & $4.5449$ & $5.7301$ \\ $10^{-2}$ & $2.99970$ & $0.332244$ & $2.98288$ & $0.334366$ & $0.5639$ & $0.6346$ \\ $10^{-3}$ & $3.00000$ & $0.333222$ & $2.99827$ & $0.333436$ & $0.0577$ & $0.0642$ \\ $10^{-4}$ & $3.00000$ & $0.333322$ & $2.99983$ & $0.333344$ & $0.0057$ & $0.0066$ \\ $10^{-5}$ & $3.00000$ & $0.333332$ & $2.99998$ & $0.333334$ & $0.0007$ & $0.0006$ \\ $10^{-6}$ & $3.00000$ & $0.333333$ & $3.00000$ & $0.333333$ & $0.0000$ & $0.0000$ \\ $10^{-7}$ & $3.00000$ & $0.333333$ & $3.00000$ & $0.333333$ & $0.0000$ & $0.0000$ \\ \hline
\end{tabular}
\caption{Comparison of the photon-sphere location and the redshift factor at the photon sphere for the braneworld wormhole. We compare the quantities obtained from the parametrized metric with those obtained from the exact metric. The last two columns give the absolute relative differences, $\delta_r=|\{(r_{\rm ph}^{\rm para}/M)-(r_{\rm ph}^{\rm exact}/M)\}/(r_{\rm ph}^{\rm exact}/M)|$ and $\delta_f=|\{f(r_{\rm ph}^{\rm para})-f(r_{\rm ph}^{\rm exact})\}/f(r_{\rm ph}^{\rm exact})|$, expressed in percentage. It can be seen that for $p\leq 10^{-6}$, these quantities start matching with that of the exact spacetime.} \label{comp-light-right-location} 
\end{table}


\begin{table}[t] \centering  \begin{tabular}{|c| c| c| c| c| c|} 
\hline $p$ & $\omega_L^{\rm para}$ & $\omega_L^{\rm exact}$ & $\delta_{\omega_R}(\%)$ & $\delta_{\omega_I}(\%)$ & $\delta_{\omega}(\%)$ \\ \hline $10^{-1}$ & $0.351038 - 0.223503\,i$ & $0.378502 - 0.241474\,i$ & $7.2560$ & $7.4422$ & $7.3103$ \\ $10^{-2}$ & $0.351558 - 0.223040\,i$ & $0.354695 - 0.225055\,i$ & $0.8844$ & $0.8953$ & $0.8876$ \\ $10^{-3}$ & $0.351942 - 0.223190\,i$ & $0.352261 - 0.223394\,i$ & $0.0906$ & $0.0913$ & $0.0908$ \\ $10^{-4}$ & $0.351985 - 0.223208\,i$ & $0.352017 - 0.223228\,i$ & $0.0091$ & $0.0090$ & $0.0091$ \\ $10^{-5}$ & $0.351989 - 0.223209\,i$ & $0.351992 - 0.223211\,i$ & $0.0009$ & $0.0009$ & $0.0009$ \\ $10^{-6}$ & $0.351989 - 0.223210\,i$ & $0.351990 - 0.223210\,i$ & $0.0003$ & $0.0000$ & $0.0002$ \\ $10^{-7}$ & $0.351989 - 0.223210\,i$ & $0.351989 - 0.223210\,i$ & $0.0000$ & $0.0000$ & $0.0000$ \\ $10^{-8}$ & $0.351989 - 0.223210\,i$ & $0.351989 - 0.223210\,i$ & $0.0000$ & $0.0000$ & $0.0000$ \\ \hline \end{tabular} \caption{Comparison of the fundamental light-ring modes of the braneworld wormhole for $\ell=1$. The parametrized and exact frequencies are compared using the same correspondence between $p$ and $\epsilon$ as in Eq.~(16). The last three columns give the absolute relative differences in the real part, in the damping rate, and in the full complex frequency, respectively: $\delta_{\omega_R}=|\{Re(\omega_L^{\rm para})-Re(\omega_L^{\rm exact})\}/Re(\omega_L^{\rm exact})|$, $\delta_{\omega_I}=\left|\{|Im(\omega_L^{\rm para})|-|Im(\omega_L^{\rm exact})|\}/|Im(\omega_L^{\rm exact})|\right|$, and $\delta_\omega=|\omega_L^{\rm para}-\omega_L^{\rm exact}|/|\omega_L^{\rm exact}|$. All three are expressed in percentage. For $p\leq 10^{-7}$ both the real and imaginary part of the $\omega_{\rm L}$ matches well with that of the exact spacetime.} \label{omwga-L-brane}\end{table}

\begin{table}[th!]
\centering
\def\arraystretch{1.25}
\setlength{\tabcolsep}{0.45em}
\resizebox{\textwidth}{!}{%
\begin{tabular}{|c||c|c|c||c|c|c||c|c|c|}
\hline
$M/a$ &
$\omega_{\rm L}^{\rm (exact)}$ &
$\omega_{\rm L}^{\rm (para)}$ &
$\omega_{\rm L}^{\rm (para)}(a_{1}=0)$ &
$\delta_{\omega_R}^{\rm para}(\%)$ &
$\delta_{\omega_I}^{\rm para}(\%)$ &
$\delta_{\omega}^{\rm para}(\%)$ &
$\delta_{\omega_R}^{a_1=0}(\%)$ &
$\delta_{\omega_I}^{a_1=0}(\%)$ &
$\delta_{\omega}^{a_1=0}(\%)$ \\
\hline
$10^{-4}$ &
$0.351964-0.223203\,i$ &
$0.351976-0.223198\,i$ &
$0.351989-0.223210\,i$ &
$3.41\times10^{-3}$ &
$2.24\times10^{-3}$ &
$3.12\times10^{-3}$ &
$7.10\times10^{-3}$ &
$3.14\times10^{-3}$ &
$6.23\times10^{-3}$ \\
\hline
$10^{-5}$ &
$0.351987-0.223209\,i$ &
$0.351988-0.223208\,i$ &
$0.351989-0.223210\,i$ &
$2.84\times10^{-4}$ &
$4.48\times10^{-4}$ &
$3.39\times10^{-4}$ &
$5.68\times10^{-4}$ &
$4.48\times10^{-4}$ &
$5.36\times10^{-4}$ \\
\hline
$10^{-6}$ &
$0.351989-0.223210\,i$ &
$0.351989-0.223209\,i$ &
$0.351989-0.223210\,i$ &
$0.00$ &
$4.48\times10^{-4}$ &
$2.40\times10^{-4}$ &
$0.00$ &
$0.00$ &
$0.00$ \\
\hline
\end{tabular}%
}
\caption{Fundamental light-ring modes of the galactic Damour-Solodukhin wormhole for $\ell=1$. We compare the light-ring modes obtained from the exact spacetime with those obtained from the parametrized spacetime. The third frequency column corresponds to a fictitious parametrized spacetime in which the leading near-field coefficient is set to $a_1=0$, included to illustrate the sensitivity of the light-ring response to the near-throat regime. In all cases, $\epsilon$ (equivalently $\lambda^2$) is kept fixed to $\lambda=10^{-3}$, while the galactic compactness $M/a$ is varied. The error columns denote absolute relative differences with respect to the exact result: $\delta_{\omega_R}=|\{Re(\omega_{\rm L}^{\rm comp})-Re(\omega_{\rm L}^{\rm exact})\}/Re(\omega_{\rm L}^{\rm exact})|$, $\delta_{\omega_I}=\left|\{|Im(\omega_{\rm L}^{\rm comp})|-|Im(\omega_{\rm L}^{\rm exact})|\}/|Im(\omega_{\rm L}^{\rm exact})|\right|$, and $\delta_{\omega}=|\omega_{\rm L}^{\rm comp}-\omega_{\rm L}^{\rm exact}|/|\omega_{\rm L}^{\rm exact}|$, where $\omega_{\rm L}^{\rm comp}$ denotes either $\omega_{\rm L}^{\rm (para)}$ or $\omega_{\rm L}^{\rm (para)}(a_1=0)$. All errors are reported in percentage.}
\label{omega-L-Gal}
\end{table}

\section{Discussion}\label{sec-9}

In this work, we constructed and analyzed a parametrized description of isolated and galactic Damour–Solodukhin and braneworld wormholes within the Bronnikov–Konoplya–Pappas framework. The central objective was to connect a systematic geometric parametrization with dynamical observables, namely electromagnetic quasinormal modes (QNMs), while ensuring consistency with shadow constraints from Sgr A*. The results clarify both the strengths and the limitations of the parametrized approach in the context of horizonless compact objects. A primary outcome of this analysis is the explicit separation between far-field and near-throat parameters. The compactified radial coordinate enables a hierarchical expansion in which the asymptotic structure is controlled independently from the strong-field geometry. 


For isolated wormholes, we demonstrated that the parametrization reproduces the qualitative structure of the effective potential, which is captured in the primary signal. At the same time, the analysis highlights an intrinsic limitation: when the exact metric functions contain non-analytical dependence on the coordinates, then the BKP parametrization cannot be used throughout the spacetime. This is a limitation of the scheme. 
The extension to a galactic Damour–Solodukhin wormhole embedded in a Hernquist halo introduces an additional scale associated with the galactic compactness. By incorporating observational bounds from the shadow of Sgr A*, we restricted the admissible parameter of our analysis.
This step is crucial: it ensures that the perturbative analysis is performed only within regions of parameter space compatible with current strong-field observations. Within this shadow-consistent regime, the resulting shifts in the QNM spectrum with $\epsilon$ are small but systematic. The oscillation frequency remains comparatively stable, reflecting the robustness of the photon-sphere structure under moderate halo contributions, whereas the damping rate exhibits greater sensitivity to the galactic compactness. 
Moreover, we have shown that even if within a certain parameter space, the metric components do not individually differ from that of the exact spacetime, the primary signal can differ if we do not include a leading near field parameter (in our case, which is $a_{1}$), which captures the response from the strong gravity regime. This indicates that the primary signal remains sensitive to the fidelity of the parametrized geometry, particularly through the light-ring location and the corresponding light-ring modes \cite{Cardoso:2008bp}.
This further demonstrates the limitation of a low-order BKP truncation when the near-throat geometry contains non-analytic structure.


It is to be stressed that, one of the important conceptual aspects of this work is its consistent linkage between three observational sectors: (i) geometric parametrization, (ii) shadow constraints, and (iii) dynamical response. The shadow measurement constrains the near photon sphere geometry; the parametrization translates those constraints into bounds on expansion coefficients; and the perturbation analysis determines how these bounds propagate into the QNM spectrum. This structured flow establishes a model-independent framework for testing wormhole geometries against strong-gravity observations. It is also worth emphasizing that electromagnetic perturbations were considered here as a clean probe of the background geometry. Unlike gravitational perturbations, they avoid complications associated with matter sector coupling and potential instabilities. Nevertheless, gravitational perturbations would be essential for direct comparison with gravitational-wave data. The present analysis therefore, provides a background for future extensions of this work to axial and polar gravitational sectors.

Several limitations of this work should also be noted at the same time. First, the study was restricted to static, spherically symmetric geometries. Rotation is expected to introduce significant qualitative changes in both shadow observables and QNM spectra. Second, only the fundamental mode was analyzed in detail; higher overtones and late-time tails may encode additional sensitivity to near-throat structure. Third, the galactic embedding was treated at the level of a static halo profile. Dynamical accretion effects and environmental perturbations were not included. Finally, while shadow constraints provide valuable bounds, future high resolution observations or GW detections could impose significantly tighter restrictions on the parameter space. Despite these limitations, the results establish that parametrized wormhole geometries can be utilized consistently with current observational data. The hierarchical parametrization used here provides a systematic way to refine the geometry, and the small deviations from exact or previously known values support the quantitative reliability of the parametrized description within its domain of validity. The combined analysis of shadow observables and quasinormal ringing therefore provides a coherent strategy for assessing the viability of horizonless compact objects as astrophysical BH mimickers.

There are several other natural directions for future work. It would be important to extend the analysis to rotating wormholes, to include gravitational perturbations, and to combine shadow and ringdown information with other observables such as lensing. Such developments would further clarify whether parametrized wormholes can remain observationally viable alternatives to classical black holes in the era of precision strong gravity measurements.

\section*{Acknowledgement}
Research of SB is supported by the NPDF fellowship provided by ANRF file no:PDF/2025/002040. SB thanks Somsubhra Ghosh for some help regarding the numerical computation of tortoise coordinate. SB wishes to thank IISER Mohali for providing hospitality during \textit{Frontiers in Gravity 2026} conference as part of this work was done there. SB also thanks IACS for providing PhD fellowship as a part of this work was done when he was a PhD student.  
SC's work is supported by MATRICS research grant awarded by Science and
Engineering Research Board (SERB), ANRF, Govt. of
India through grant no. MTR/2022/000318. The authors thank the unknown referee for making valuable comments.
\appendix
\section{Problem of parametrizing the galactic wormhole}\label{app-1}
In \ref{sec-6}, we stated that the BKP scheme can be used to parametrize the $g_{rr}$ component of the metric. But due to transcendental nature of $g_{tt}$, we have not directly used the scheme. Here we will use the BKP parametrization directly to obtain the values of the parameters and check if the obtained values make sense. For simplicity, we define the followings:
\begin{align}\label{Simple-Symbols}
   & \delta\equiv\frac{a-M}{\sqrt{M(2a-M+4M_{1})}}~,\\
    &\xi=\sqrt{M(2a-M+4M_{1})}~,\\
    &\chi=\sqrt{\frac{M}{2a+4M_{1}-M}}~.
\end{align}
So that,
$e^{\gamma(r)}$ becomes,
\begin{align}
    e^{\gamma(r)}=e^{-\chi \pi}e^{2\chi \tan^{-1}\left(\delta+\frac{r}{\xi}\right)}~.
\end{align}
Now at large $r$, it behaves as follows:
\begin{align}
   e^{\gamma(r)}\approx \left(1-\frac{2\xi \chi}{r}+\frac{2\chi\xi^{2}(\delta+\chi))}{r^{2}}+O\left(\frac{1}{r^{3}}\right)\right)~.
\end{align}
So that the large $r$ behavior of $f(r)$ becomes,
\begin{align}\label{f-large}
   f(r)\approx 1-\frac{2}{r}\left(\xi\chi+M_{1}\right)+\frac{2}{r^{2}}\left[\chi\xi^{2}(\delta+\chi)+2M_{1}\xi\chi\right]+O\left(\frac{1}{r^{3}}\right)~.
\end{align}
Then upon simplification and using the relation between $x$ and $r$ we get the following expression,
\begin{align}\label{f-gal-exact-far-x}
   f\approx 1-\frac{2(M+M_{1})}{2M_{2}}(1-x)+\frac{(a+2M_{1})M}{2M_{2}^{2}}(1-x)^{2}+O\left((1-x)^{3}\right)~.
\end{align}
Comparing this expression with~~\ref{a0}~~ we get the far field parameters,
\begin{align}\label{far-gal-exact}
    \epsilon=\frac{M+(M_{1}-M_{2})}{M_{2}}~,\\
    a_{0}=\frac{M(a+2M_{1})}{2M_{2}^{2}}~.
\end{align}
These expressions appear reasonable at first sight, but invoking into physical scenarios, one makes the assumption that $a\gg M\gg M_{1}$. However, in that case, one can easily check that the condition~~\ref{hierarchy-parameters} will be violated and the hierarchy of the parameters breaks down.\\
\\
Next, to get the small $x$ behavior of $e^{\gamma(r)}$, we write it as,
\begin{align}\label{e-gamma-x}
   e^{\gamma}=e^{-\chi \pi}e^{2\chi\tan^{-1} \left(\delta+\frac{2M_{2}}{\xi(1-x)}\right)}~. 
\end{align}
From the first sight, it will appear that the expression is non-analytic in $x$. However, in our case such non-analyticity is harmless in the following sense that if $x\rightarrow 1^{-}$, then $\tan^{-1}(\frac{1}{1-x})=+\frac{\pi}{2}$. Strictly speaking, the two-sided limit $x\rightarrow 1$ does not exists as for $x\rightarrow 1^{+}$, it will give $-\frac{\pi}{2}$. However since $x\in [0,1]$,~so that our interest is in the former case.\\
\\
\smallskip
Now for small values of $x$, that is, for $x\rightarrow 0$ ($r \rightarrow 2M_{2}$) we get
\begin{align}\label{e-gamma-small-x-1}
    e^{\gamma(r)}\approx e^{-\chi \pi}e^{2\chi \tan^{-1}\left(\delta+\frac{2M_{2}}{\xi}\right)}\left[1+\frac{4M_{2}M}{(2M_{2}+a-M)^{2}+M(2a-M+4M_{1})}~x~+O(x^{2})\right]~.
\end{align}
This leads to the following small $x$ behavior of $f(r)$, namely,
\begin{align}\label{f-samll-x-behavior}
    f\approx  e^{-\chi \pi}e^{2\chi \tan^{-1}\left(\delta+\frac{2M_{2}}{\xi}\right)}\left[1-\frac{M_{1}}{M_{2}}~+~\frac{M_{1}}{M_{2}}x\right]\left[1+\frac{4M_{2}M}{(2M_{2}+a-M)^{2}+M(2a-M+4M_{1})}~x~+O(x^{2})\right]~.
\end{align}
One can easily check that because of the exponential prefactor, the expansion of $f$ in terms of $x$ will in general not match with \ref{A-o(x)}. Therefore, as claimed in the text that the transcendental nature of $g_{tt}$ limits the direct applicability of BKP scheme. One may argue that rescaling $f(r)$ by the prefactor will resolve the issue but in that case the prefactor will popup in the expansion \ref{f-gal-exact-far-x} and will lead to a similar problem.\\
\\
One can still proceed further with the assumption that $a\gg M\gg M_{1}$,~~which leads to the following approximate expression,
\begin{align}\label{e-gamma-small-x-2}
     e^{\gamma(r)}\approx \left[1+\frac{4M_{2}M}{(2M_{2}+a-M)^{2}+M(2a-M+4M_{1})}~x~+O(x^{2})\right]~.
\end{align}
Therefore, the small $x$ behavior of $f(r)$ turns out to be:
\begin{align}\label{f-gal-small-x}
    f\approx \left(1-\frac{M_{1}}{M_{2}}\right)+\left[\frac{4M (M_{2}-M_{1})}{(2M_{2}+a-M)^{2}+M(2a-M+4M_{1})}+\frac{M_{1}}{M_{2}}\right]~x~+O(x^{2}).
\end{align}
Comparing the above expression with~~\ref{A-o(x)}~and the values of $\epsilon$ and $a_{0}$, we get,
\begin{align}\label{fo and a1}
    &f_{0}=\left(1-\frac{M_{1}}{M_{2}}\right)~,\\
    &a_{1}=\frac{2M}{M_{2}}-\frac{M(2a+M_{1})}{2M^{2}_{2}}+\frac{4 M(M_{2}-M_{1})}{(a+2M_{2})^{2}+4M(M_{1}-M_{2})}~.
\end{align}
Moreover, using the parameters \ref{fo and a1} and \ref{far-gal-exact} in \ref{A}, one can see that it leads to the unphysical behavior, namely for some values of $x$, $A(x)$ becomes larger than $1$ . For these issues, we have avoided the direct use of the BKP scheme and used the approximate parametrization as discussed in the text. Moreover, one can check that for $M=0$, these parameters reduce to that of the isolated DS wormhole (see~\ref{sec-3}).

\bibliography{biblio}
\bibliographystyle{utphys1}
\end{document}